\documentclass[superscriptaddress,aps,preprintnumbers,amsmath,amssymb,prd,nofootinbib,preprint]{revtex4}
\usepackage{bm,color,xcolor} 
\usepackage{amssymb,amsfonts,slashed,amsthm,amsmath,graphicx,soul,physics}
\usepackage{epsfig,subfigure}
\usepackage{ulem}
\usepackage{multirow,array}

\begin{document}


\preprint{TU-1215}

\title{
Stability of domain walls with inflationary fluctuations under potential bias, and  gravitational wave signatures
}

\author{
Naoya Kitajima
}

\affiliation{Frontier Research Institute for Interdisciplinary Sciences, Tohoku University, Sendai, Miyagi 980-8578, Japan}
\affiliation{Department of Physics, Tohoku University, 
Sendai, Miyagi 980-8578, Japan}

\author{
Junseok Lee
}

\affiliation{Department of Physics, Tohoku University, 
Sendai, Miyagi 980-8578, Japan}

\author{
Fuminobu Takahashi
}

\affiliation{Department of Physics, Tohoku University, 
Sendai, Miyagi 980-8578, Japan}

\author{
Wen Yin
}

\affiliation{Department of Physics, Tohoku University, 
Sendai, Miyagi 980-8578, Japan}

\begin{abstract}
A recent study has shown that domain walls with  inflationary initial fluctuations exhibit remarkable stability against population bias due to long-range correlations, challenging the claims of prior research. In this paper, we study the dynamics of these domain walls in the presence of potential bias and show that they collapse with a lifetime several times longer than that due to thermal fluctuations. This is interpreted as a difference in the average distance between domain walls,
leading us to derive a new formula for the domain wall lifetime which depends on the area parameter in a qualitatively different way from previous studies.
In addition, we compute the spectrum of gravitational waves generated by such domain walls and find that both the peak frequency and the peak abundance are lowered in a manner that depends on the area parameter. 
Based on these findings, we also determine the necessary degree of vacuum degeneracy for axion domain walls to explain the isotropic cosmic birefringence.

\end{abstract}

\maketitle
\flushbottom

\vspace{1cm}

\section{Introduction}
Spontaneous symmetry breaking (SSB) leads to various topological defects that inherit information from high-energy physics. While the stability of individual defects is ensured by their topological nature, the stability and evolution of networks of defects are not always obvious.  By studying the cosmological evolution of their network, we can learn about physics beyond the Standard Model and the evolution of the early universe. In this paper, we examine the stability of domain walls, which result from SSB of discrete $Z_2$ symmetry~\cite{Zeldovich:1974uw,Kibble:1976sj,Vilenkin:1981zs} (see Ref.~\cite{Vilenkin:2000jqa} for reviews).

The simplest model that produces domain walls is the $Z_2$ model given by
\begin{equation}
V_0(\phi) = \frac{\lambda}{4} (\phi^2-v^2)^2
= \frac{\lambda}{4} v^4 - \frac{1}{2} m_0^2 \phi^2 + \frac{\lambda}{4} \phi^4
\label{eq:pot}
\end{equation}
with $m_0 \equiv \sqrt{\lambda} v$. 
Here $\phi$ is a real scalar field, $\lambda (> 0)$ is the quartic coupling, and there are two degenerate vacua located at $\phi = \pm v$ due to the $Z_2$ symmetry, $\phi \to - \phi$.
 Domain walls are structures that separate these two vacua, and their tension $\sigma$ is given by $\sigma \sim m_0 v^2$. 
 Domain walls are known to quickly reach the scaling solution, where there is roughly one domain wall per Hubble horizon~\cite{Press:1989yh,Hindmarsh:1996xv,Garagounis:2002kt,Oliveira:2004he,Avelino:2005kn,Leite:2011sc,Leite:2012vn,Martins:2016ois}. The energy density of domain walls following the scaling solution decreases more slowly than non-relativistic matter or radiation, and therefore eventually could dominate the universe. Then, domain walls would cause too large anisotropies in the cosmic microwave background radiation (CMB) in contradiction with observations, unless their tension satisfies the Zel’dovich bound, $\sigma \lesssim ({\rm MeV})^3$~\cite{Zeldovich:1974uw}. This is the cosmological domain wall problem~\cite{Zeldovich:1974uw,Vilenkin:1984ib}.

Several mechanisms have been suggested to destabilize domain walls to avoid this problem. In particular, introducing bias in the potential and/or in the initial distribution has been  investigated in the literature. When a bias is introduced in the potential, domain walls collapse soon after the false vacuum pressure becomes stronger than their tension~\cite{Gelmini:1988sf,Larsson:1996sp,Lalak:2007rs,Correia:2014kqa,Correia:2018tty}. On the other hand, when bias is introduced in the initial distribution, which we call population bias, some studies claimed that domain walls disappear quickly for both thermal and initial inflationary fluctuations~\cite{Lalak:1993ay,Lalak:1993bp,Lalak:1994qt,Coulson:1995uq,Coulson:1995nv,Larsson:1996sp,Lalak:2007rs,Krajewski:2021jje}. This was a common belief in this field. However, Gonzalez and three of the present authors (NK, FT, and WY) recently showed that this common belief is far from reality and that domain walls with inflationary fluctuations as initial conditions are very stable against the population bias~\cite{Gonzalez:2022mcx}.\footnote{
The error in the past literature arose from improperly applying percolation theory to fluctuations with superhorizon scale correlations, combined with a failure in the numerical generation of inflationary fluctuations for the initial conditions.
} The reason for this stability is that  correlations at superhorizon scales are preserved since the dynamics of the domain walls is locally governed by the random motion in the absence of potential bias. The purpose of this paper is to clarify how the stability of domain walls with initial inflationary fluctuations changes with potential bias.

Axion domain walls are an example of such domain walls with inflationary fluctuations as initial conditions~\cite{Linde:1990yj} (see also Refs.\,\cite{Lyth:1991ub,Nagasawa:1991zr,Khlopov:2004sc,Davoudiasl:2006ax,Kitajima:2015nla,Daido:2015gqa,Daido:2015bva,Kobayashi:2016qld,Liu:2020mru, Sakharov:2021dim,Takahashi:2020tqv,Kitajima:2022jzz,Gonzalez:2022mcx,Redi:2022llj,Gorghetto:2023vqu}). Axions are known to have a periodic potential and thus have domain wall solutions separating degenerate minima. Since the axion potential is stable against perturbative radiative corrections, axions naturally remain light during inflation and acquire quantum fluctuations.  The recent finding of Ref.~\cite{Gonzalez:2022mcx} implies that axions with inflationary fluctuations, which are naturally biased towards one of the vacua, could play an important cosmological role through their domain walls. 

One of the interesting cosmological implications of domain walls is the generation of gravitational waves from their collapse~\cite{Vilenkin:1981zs,Preskill:1991kd,Chang:1998tb,Gleiser:1998na,Hiramatsu:2010yz,Kawasaki:2011vv,Hiramatsu:2013qaa,Higaki:2016jjh,Nakayama:2016gxi,Saikawa:2017hiv,Ferreira:2022zzo,Kitajima:2023cek}. 
Recently, Murai and the present authors have for the first time calculated gravitational waves generated during the collapse of domain walls~\cite{Kitajima:2023cek} and have shown that the axion domain walls coupled to QCD naturally explain the stochastic gravitational waves in the nHz range recently detected by the pulsar timing arrays~\cite{NANOGrav:2023hvm,Antoniadis:2023ott,Reardon:2023gzh,Xu:2023wog} (see also Refs.~\cite{Higaki:2016jjh,Ferreira:2022zzo,Blasi:2023sej} for the generation of nHz gravitational waves from axion domain walls coupled to QCD).
Later in this paper, we will estimate the gravitational wave spectrum for the two initial conditions, thermal and inflationary fluctuations. Especially the latter is calculated for the first time in this paper.
Another is the cosmic birefringence due to axion domain walls coupled to photons~\cite{Takahashi:2020tqv,Kitajima:2022jzz,Gonzalez:2022mcx,Kitajima:2023cek}; they can nicely explain the close proximity of the observed isotropic rotation angle of the CMB polarization~\cite{Minami:2020odp,Diego-Palazuelos:2022dsq} to the fine structure constant, and also predict a characteristic power spectrum of the anisotropic cosmic birefringence within the reach of future CMB observations.

In this paper, we study how a network of domain walls with initial inflationary fluctuations  evolves under a potential bias. We find that while these domain walls are remarkably stable against population bias, they decay under potential bias with a lifetime a few times longer than those with thermal fluctuations. 
We derive a formula for the domain wall lifetime which depends on the area parameter in a qualitatively different way from previous studies.
We then calculate the spectrum of gravitational waves generated in the scaling regime of the domain walls for the two initial conditions, 
and show that the difference can be attributed to the area parameter for the two cases.
We also derive an upper limit on the energy difference between adjacent vacua to account for cosmic birefringence with axion domain walls.

The rest of this paper is organized as follows. In Sec.~\ref{sec2}, we explain the set-up with the domain wall solution and the two representative initial conditions. In Sec.~\ref{sec3} we introduce the two types of biases, population bias and potential bias. Our numerical results are presented in Sec.~\ref{sec4}. The gravitational wave spectrum is calculated in Sec.~\ref{sec5}. The last section is devoted to discussion and conclusions.

\section{Domain-wall network}
\label{sec2}
\subsection{Set-up}
The Lagrangian for a real scalar field $\phi(x)$ is given by
\begin{align}
    \mathcal{L} = -\frac{1}{2}\partial_{\mu} \phi \hspace{0.3mm}\partial^{\mu} \phi - V(\phi),
    \label{eq:Lagr}
\end{align}
where  $V(\phi)$ is the potential, and we consider $V(\phi) = V_0(\phi)$ given in Eq.~(\ref{eq:pot}) for the moment. Later we will add an extra term that explicitly breaks the $Z_2$ symmetry.
In the Friedmann-Lemaître-Robertson-Walker (FLRW) universe, the equation of motion of  $\phi$ is given by
\begin{equation}
\ddot{\phi} +3H\dot{\phi} -\frac{1}{a^2}\Delta\phi +\frac{\partial V}{\partial\phi} = 0,
\label{eq:EoM}
\end{equation}
where $H$ is the Hubble parameter, the dot denotes the time derivative, and $\Delta$ is the Laplacian with respect to the comoving coordinate.

Throughout this paper, we consider the radiation-dominated universe with $H=1/2t$.
For later use let us define the conformal Hubble parameter ${\cal H}$ and the conformal time $\tau$;
\begin{align}
{\cal H} &\;\equiv\; a H,\\
\tau & \;\equiv \; \int_0^t \frac{dt'}{a(t')} = \sqrt{\frac{2t}{m_0}},
\end{align}
where we set the scale factor equal to unity when $H = m_0$ in the last equality. With this normalization, the scale factor is given by $a(\tau) = m_0 \tau$.


We assume that the scalar field $\phi$ initially remains near the local maximum at the origin, accompanied by some fluctuations. In the $Z_2$ model, this assumption is reasonable, since the origin is a symmetry-enhanced point and can be stabilized by additional corrections such as thermal mass in the very early universe. If the field is in thermal contact with the plasma, the initial fluctuation is due to thermal fluctuations. On the other hand, if the scalar remains light during inflation and is decoupled after inflation, the fluctuation is inflationary with correlation at superhorizon scales.

Domain walls are characterized by their tension. For the $Z_2$ potential $V_0(\phi)$, the tension $\sigma$ is given by
\begin{equation}
\sigma = \frac{2\sqrt{2}}{3}\sqrt{\lambda} v^3 = \frac{2\sqrt{2}}{3}\frac{{m_0}^3}{\lambda}.
\end{equation}
The width of domain walls is of order $m_0^{-1}$, and at scales larger than the width,
the properties of domain walls do not depend on details of the potential shape and are characterized only by the tension.
Thus, we will be able to apply our results based on the $Z_2$ model to the sine-Gordon potential for the axion later in this paper.


\subsection{Scaling solution}
The energy of the domain-wall network is approximately given by the product of its area and the tension, if it is non-relativistic. Thus, once formed, domain walls move around to minimize their surface area due to their tension. When domain walls collide, they disappear, further reducing the surface area. Since the motion of domain walls tends to minimize the area only locally, the typical radius of curvature cannot significantly exceed the Hubble horizon scale. As a result, the domain-wall network is known to approach the situation where approximately one domain wall is contained within each Hubble volume. This is the so-called scaling solution, for which the (spatially averaged) energy density of the domain walls, $\rho_{\rm DW}$, is given by
\begin{align}
\rho_{\rm DW} \sim \sigma H.
\end{align}
The scaling solution has been numerically confirmed in the literature~\cite{Press:1989yh,Hindmarsh:1996xv,Garagounis:2002kt,Oliveira:2004he,Avelino:2005kn,Leite:2011sc,Leite:2012vn,Martins:2016ois}. 
However, most of the literature investigates it for white noise initial conditions corresponding to thermal fluctuations. On the other hand, Ref.~\cite{Gonzalez:2022mcx} is the first to correctly investigate the scaling solution for inflationary fluctuations, which is discussed below.

In three-dimensional space, the scaling solution can also be expressed in terms of the averaged area per Hubble horizon,  $A_{\text{DW}}$, as
\begin{equation}
   \mathcal{A}_3 \equiv  \frac{A_{\text{DW}}}{H^{-2}} = \mathcal{O}(1).
    \label{eq:ScalingSol3}
\end{equation}
Later we will numerically study the case with translation symmetry along one of the spatial directions, e.g. the $z$-axis, where the domain walls appear as lines in the remaining two-dimensional space. In this case, the scaling solution corresponds to
\begin{equation}
   \mathcal{A}_2 \equiv \frac{L_{\text{DW}}}{H^{-1}} = \mathcal{O}(1),
    \label{eq:ScalingSol2}
\end{equation}
where $L_{\text{DW}}$ is the average length of the domain wall contained within a Hubble horizon.
We call $\mathcal{A}_2$ and $\mathcal{A}_3$ as the area parameter. Note that the area parameter defined here differs
from the one often defined in the literature  as $\rho_{\rm DW} = \frac{\mathcal{A}\sigma}{t}$~\cite{Hiramatsu:2013qaa}. When the motion of domain walls is non-relativistic, we have $\mathcal{A}_3 \approx 2 \mathcal{A}$ in the radiation-dominated universe. Note that ${\cal A}_3$ is defined in terms of area, while ${\cal A}$ is defined using energy density.

The exact value of the $\mathcal{O}(1)$ constant on the RHS of the above equations depends on the initial conditions~\cite{Gonzalez:2022mcx}. For the two initial conditions we introduce in the next subsection, they differ by a factor of about $2$.
This difference actually comes from the large-scale correlation of the domain-wall network encoded by the initial condition.\footnote{
This statement is based on the assumption that the deviation of the area parameter observed in our simulations is a genuine physical effect rather than a transient one. Although our domain wall network simulations are performed with fairly high precision, the area parameter still shows a mild time dependence for both initial conditions, and full convergence has not yet been demonstrated. See  Appendix \ref{app:2}. Confirming full convergence would require simulations with a much larger dynamical range, which is challenging at present.}
We emphasize here that the energy density or surface area (or length) is a spatially averaged quantity and does not capture well the global properties of the domain-wall network. We need to compute the power spectrum of the scalar field to directly see the global structure of the domain-wall network.

\subsection{Initial conditions}

Scalar field fluctuations can be generated from a variety of sources, but typical ones are thermal and inflationary fluctuations. In this paper, we consider these two initial fluctuations, but it is straightforward to perform other types of fluctuations. 

To represent the initial fluctuations, it is convenient to use their two-point correlation function or power spectrum;
\begin{equation}
\label{eq:psphi}
\left\langle{\phi(\boldsymbol{k})\phi(\boldsymbol{k'})}\right\rangle = (2\pi)^d\delta^{(d)}(\boldsymbol{k}+\boldsymbol{k'})P(k),
\end{equation}
where $\phi({\boldsymbol k}) = \int d^d {\boldsymbol x}\, e^{- i {\boldsymbol k}\cdot{\boldsymbol x}} \phi({\boldsymbol x})$ is the Fourier mode of $\phi({\boldsymbol x})$, $d = 2, 3$ is the spatial dimension, and $P(k)$ is the power spectrum {with $k$ being $|{\boldsymbol k}|$}. Here, we have assumed statistical isotropy, which implies that the power spectrum depends solely on the momentum modulus. For the sake of brevity, we have omitted the explicit time dependence in our notation.

The variance of fluctuations in real space is related to the power spectrum as
\begin{equation}
\left\langle{\phi^2(\boldsymbol{x})}\right\rangle 
= \int \frac{d^d \boldsymbol{k}}{(2\pi )^d} P(k)
= \int d \ln k \,\mathcal{P}(k),
\end{equation}
where $\mathcal{P}(k)$ is the reduced power spectrum, and it is given by $\mathcal{P}(k) = k^2P(k)/2\pi$ for $d=2$, and $\mathcal{P}(k) = k^3P(k)/2\pi^2$ for $d=3$.

The vacuum fluctuation of a scalar field with mass $m$ in the flat Minkowski space
is given by
\begin{equation}
    \left\langle{\phi(\boldsymbol{k})\phi(\boldsymbol{k'})}\right\rangle = \frac{1}{2\sqrt{k^2+m^2}}\delta^{(d)}(\boldsymbol{k}+\boldsymbol{k'})
\end{equation}
with some ultraviolet cutoff. On scales larger than $m^{-1}$, the power spectrum is independent of ${\boldsymbol k}$. This is the same as white noise, which is defined to have a constant power spectrum, $P(k) = {\rm const.}$, or equivalently, ${\cal P}(k) \propto k^d$. In real space, white noise is often implemented by generating the random field values at each lattice site. Similar fluctuations are obtained for thermal fluctuations on scales larger than the thermal correlation length. The white noise fluctuation is a plausible choice for the initial condition if there were no correlations at superhorizon scales because, in that case, the field value averaged over the Hubble horizon should follow the random distribution. In most of the literature, the white noise initial condition has been adopted in the numerical lattice calculations of the topological defects. This choice of the initial fluctuation is in line with the spirit of the Kibble mechanism~\cite{Kibble:1976sj}. In the following, thermal fluctuation and white noise are used interchangeably.

On the other hand, the initial fluctuations at superhorizon scales could have non-trivial correlations. This is the case if the corresponding scalar is light during inflation. Then, it acquires quantum fluctuation of order the Hubble parameter during inflation, $H_{\rm inf}$, leading to the (almost) scale-invariant power spectrum, 
\begin{align}
    \mathcal{P}(k) = \left(\frac{H_{\rm inf}}{2\pi}\right)^2
\end{align}
at $k \lesssim aH = \mathcal{H}$. Compared to the white noise fluctuations, it is not suppressed at large scales, implying that the domain-wall network has correlations at superhorizon scales. This initial inflationary fluctuation was correctly implemented for the first time in numerical lattice simulations of domain walls in Ref.~\cite{Gonzalez:2022mcx}, where it was found that the structure of the domain-wall network as well as its subsequent evolution is completely different from that with the white noise initial condition. In the following, inflationary fluctuation refers to this scale-invariant fluctuation. 

\begin{figure}[!t]
    \begin{center}  
        \includegraphics[width=160mm]{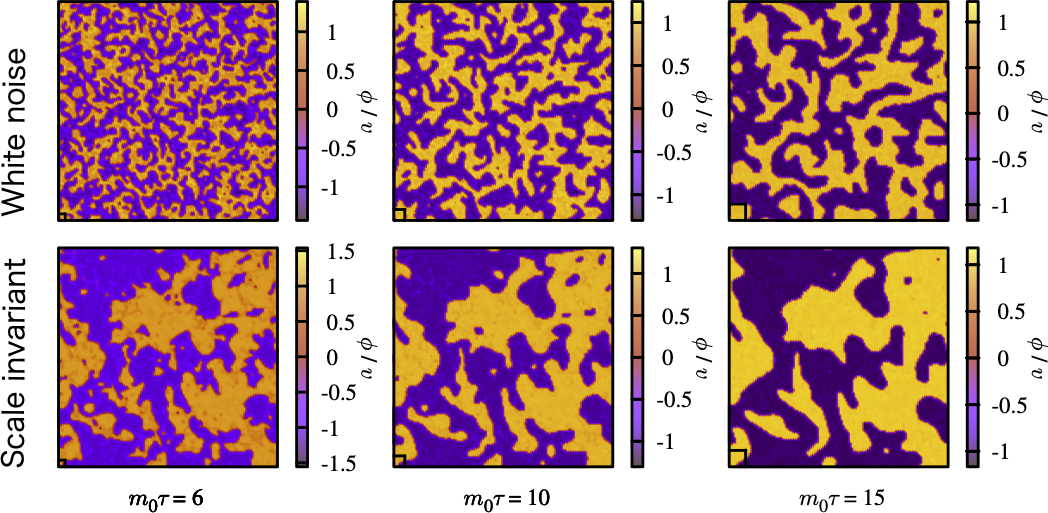}
    \end{center}
    \caption{
    The snapshots of the domain wall distributions at different times $\tau = 6, 10, 15/m_0$ from left to right, where the initial conditions are provided by either the white noise (top) or the scale-invariant fluctuations (bottom). 
    No bias is introduced. Approximately $67^2, 20^2, 13^2$ Hubble horizons are contained in order of increasing time. The size of the Hubble horizon is indicated by the small box in the lower left of each plot. The simulations are performed on an $8192^2$ comoving lattice. The color scheme represents the field value normalized by $v$. 
    }
    \label{fig:realspace_snapshot_unbiased}
\end{figure}

In Fig.~\ref{fig:realspace_snapshot_unbiased} we show snapshots of the domain-wall network at different times for the white noise (top) and the inflationary fluctuations (bottom). It is evident that the global network structure differs significantly between these cases. In particular, in the case of inflationary fluctuations, the global structure of the domain-wall network remains preserved as it undergoes coarse-graining over time. The persistent stability of this global structure serves as the underlying reason for the robustness of domain walls with inflationary fluctuations~\cite{Gonzalez:2022mcx}.

\section{Destabilizing domain walls with biases}
\label{sec3}

The stability of domain walls is ensured by their topological nature and, in the case of the $Z_2$ model, by the $Z_2$ symmetry. Since they follow the scaling solution, their energy density decreases more slowly than that of radiation or non-relativistic matter. Consequently, domain walls eventually come to dominate the energy density of the universe, leading to an unwanted inhomogeneity that contradicts observations. This is known as the cosmological domain wall problem~\cite{Zeldovich:1974uw}. To avoid this problem, there are two possibilities: either the domain wall tension must be less than $\sim ({\rm MeV})^3$, or the domain-wall network must become unstable and decay before reaching dominance.

\begin{figure}[!t]
    \begin{center}  
        \includegraphics[width=160mm]{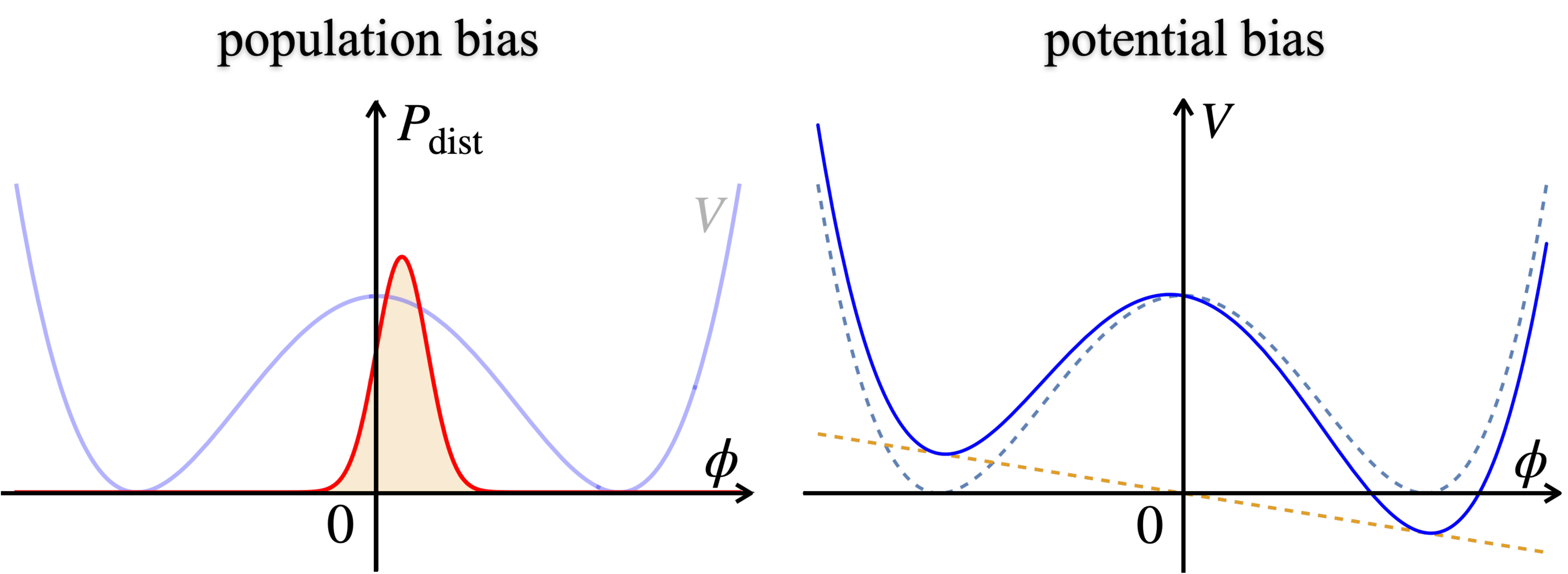}
    \end{center}
    \caption{
    Illustration of the two biases introduced to destabilize the domain-wall network. (Left) Population bias: The initial distribution of the scalar field  is biased toward one of the vacua, resulting in an uneven population of vacua. (Right) Potential bias: An explicit $Z_2$ breaking term lifts the degeneracy of the two vacua, leading to the preferential motion of the domain walls. }
    \label{fig:asymmetry}
\end{figure}

In this study, we explore the latter possibility by considering the breaking of the $Z_2$ symmetry. Two biases can be introduced to render the domain-wall network unstable: the initial population bias and the potential bias, as shown in Fig.~\ref{fig:asymmetry}. The left panel shows the initial population bias, which corresponds to a biased distribution of the scalar field in the initial state. As a result, one vacuum occupies a larger or smaller volume than the other, causing the domain walls to move in a way that amplifies the bias over time due to their tension. On the other hand, the right panel shows the potential bias, which affects the dynamics of the domain walls in a more intuitive way. The potential bias induces negative pressure in the false vacuum, causing the domain walls to move toward it. When the pressure acting on the walls exceeds their tension, the domain walls progressively converge toward the false vacuum. Thus, for the population bias, the local behavior of the domain walls is the same as under unbiased conditions. In contrast, the potential bias introduces a preferred direction of motion that affects the domain wall dynamics on both local and global scales.

The effects of both biases on the domain-wall network have been extensively studied in the literature for two different initial conditions: white noise and inflationary fluctuations. Previous studies have shown that both biases contribute to the rapid decay of the domain-wall network, with the initial population bias having a particularly strong impact in both cases~\cite{Lalak:1993ay,Lalak:1993bp,Lalak:1994qt,Coulson:1995uq,Coulson:1995nv,Larsson:1996sp,Lalak:2007rs,Krajewski:2021jje}. However, it is crucial to note that previous numerical simulations did not properly incorporate the initial inflationary fluctuations, resulting in a substantial underestimation of  correlations at the superhorizon scales. Also, the use of percolation theory in the past literature to analyze inflationary fluctuations has been misguided due to the overlooked long-range correlations, which fundamentally invalidate its application.

A recent study by Gonzalez and some of the present authors~\cite{Gonzalez:2022mcx} has shed new light on this issue. Surprisingly, the domain-wall network with initial inflationary fluctuations was found to be remarkably stable to the initial population bias, contrary to the prevailing understanding in the community. This unexpected robustness can be attributed to the presence of correlations at superhorizon scales, as shown in Fig.~\ref{fig:realspace_snapshot_unbiased}. Motivated by these findings, the main objective of this paper is to investigate the stability of the domain-wall network with the initial inflationary fluctuations in the presence of the potential bias.

As a specific example of the potential bias, we introduce a linear term;
\begin{align}
    V(\phi) &\;=\; V_0(\phi) + \delta V(\phi), \\
    \delta V(\phi) &\;=\; -\epsilon\lambda v^3\phi \quad ,
    \label{eq:BiasTerm} 
\end{align}
where $V_0(\phi)$ respects the $Z_2$ symmetry, while the linear term $\delta V$ explicitly breaks the $Z_2$ symmetry. Here $\epsilon$ denotes a dimensionless parameter that controls the magnitude of the potential bias, and we assume
\begin{equation}
    \epsilon \geq 0
\end{equation}
without loss of generality.

The linear term $\delta V$ induces a pressure on the wall toward the false vacuum,
and it is approximately given by the potential difference between two local minima $\Delta V$:
\begin{equation}
    p \approx \Delta V \approx 2 \epsilon \lambda v^4 = \frac{2 \epsilon m_0^4}{\lambda}.
    \label{eq:pressure} 
\end{equation}
This can be derived by taking the time derivative of the momentum of the wall, which is estimated by the integral of the stress-energy tensor along a line 
perpendicular to the wall.
This pressure $p$ affects the dynamics of the domain wall as it competes with the tension due to the curvature of the wall.
When the domain walls are in the scaling regime, the force due to tension can be approximated by  $\sigma /R \sim \sigma H$, where $R\sim H^{-1}$ is
a typical curvature radius.

Lastly, let us comment on a subtle point. The introduction of a linear term as a potential bias can shift the position of the local maximum away from the origin. Consequently, if we were to consider initial fluctuations around the origin in the presence of a linear term, we would be introducing a population bias as well. Since our primary focus is to examine the impact of potential bias on the evolution of the domain-wall network, we will consider fluctuations centered around the local maximum in our numerical calculations. This approach aims to minimize the effect of any secondary population bias. In the present set-up, the position of the local maximum $\phi_{\text{max}}$ is given by
\begin{align}
    \phi_{\text{max}} \simeq -(\epsilon+\epsilon^3+3\epsilon^5)v,
    \label{eq:phi_max}
\end{align}
and we ensure that the ensemble average of the initial scalar field values is equal to
$\phi_{\text{max}}$.

\section{Lifetime of domain walls under biases}
\label{sec4}

\subsection{Set-up for numerical simulations}
\label{subsec4A}

We have numerically solved the equation of motion (\ref{eq:EoM}) on the two-dimensional $N_{\rm grid}^2$ lattice with $N_{\rm grid} = 8192$ or $16384$, using the leap-frog method.
The comoving size of lattice box $L_{\rm box}^2$ is set to be $L_{\rm box}^2 = (64\pi/m_0)^2$.
All simulations start at the initial conformal time $\tau_i =1/m_0$ where $H$ is equal to $m_0$.
The parameters we used in simulations are shown in Table~\ref{tab:params}. 

As explained in Sec.~\ref{sec2}, we compare the cases of two types of initial fluctuations, the white noise and the scale-invariant fluctuations. For the numerical calculations, we choose the momentum space to set the initial distributions, because it is easy to handle the wavenumber dependence. The white noise fluctuation is generated by assigning the Gaussian distributed random number to each discretized momentum lattice site. On the other hand, the scale-invariant fluctuation can be generated by assigning the Gaussian distributed random number weighted by some power of the comoving wavenumber $k$ on the momentum space, i.e. $k^{-1}$ for 2D and $k^{-3/2}$ for 3D.
For both cases, we introduce the ultraviolet cutoff $k_{\rm UV}$ as the Hubble scale, $k_{\mathcal{H}} = 2\pi \mathcal{H}$, for the initial condition to suppress the gradient energy and make the system easy to get to the scaling solution. Note that the fluctuations are normalized in different ways. In the white noise case, the initial fluctuations are normalized so that the standard deviation in the real space, $\sqrt{\langle (\phi - \langle \phi \rangle)^2 \rangle}$,  is equal to $0.1v$. On the other hand, we set $\sqrt{\mathcal{P}(k)}$ to $0.1v$ in the case of scale-invariant fluctuation.\footnote{In the real space, this corresponds to $\sqrt{\langle (\phi - \langle \phi \rangle)^2 \rangle} \simeq 0.23v$.}
This is because the fluctuations are accumulated in real space for the scale-invariant fluctuation, and the variation in real space depends logarithmically on the box size. The initial conditions of Fig.~\ref{fig:realspace_snapshot_unbiased} are generated this way.
To neglect the contribution from the population bias, we set the average field value at $\phi = \phi_{\rm max}$. This setup can be also realized in natural models (see Appendix.~\ref{app:1}).

Let us discuss some subtleties about the field variance in real space in the case of inflationary fluctuations. As previously mentioned, the variance increases logarithmically as the box size expands. Once the integral $\int d \ln k\, \mathcal{P}(k) \gtrsim v^2$, the quartic term becomes important in the $Z_2$ model, and at some point, the scalar is no longer sufficiently light. This constrains the range of the scalar fluctuations to a finite range. Let us consider what happens if the field value of $\phi$ is much larger than $v$, but its effective mass squared $\sim V''(\phi)$ is still smaller than the Hubble parameter. In such a case we can adopt the inflationary fluctuations. When the Hubble parameter becomes comparable to the effective mass after inflation, $\phi$ starts to oscillate with an amplitude much larger than $v$ and can fall into either a positive or negative vacuum. Thus, as the size of the box increases, positive and negative vacua are randomly populated at sufficiently large scales. Consequently, the statistical properties of the fluctuations gradually transition from inflationary to thermal as one moves to larger scales. This transition can be interpreted as the restoration of the $Z_2$ symmetry due to the effective mass arising from self-coupling, which only becomes important at large scales. In this limit, the resulting domain-wall network with inflationary fluctuations is realized only in rare regions of the large box.  We have confirmed this transition numerically.  However, it occurs only when considering a case of $\int d \ln k\, \mathcal{P}(k) \gtrsim v^2$, and it also depends on the potential shape beyond the potential minima. In fact, such a transition does not occur in the case of the cosine potential, and other vacua are populated. In this case, the field correlation remains on a large scale. In the following, we focus on a regime where the variance of the scalar field is sufficiently small so that the evolution of the domain-wall network does not depend on the potential shape beyond the potential minima.\footnote{
This transition can be used to suppress isocurvature perturbations of  $\phi$ particles. As we will discuss later, if $\phi$ particles produced from the collapse of the domain walls with inflationary fluctuations contribute  to dark matter, isocurvature perturbations may become too large. However, by considering the transition to white noise at large scales discussed here, the isocurvature perturbations of $\phi$ can be suppressed.
}

In simulations of domain walls, two distinct physical scales emerge: the Hubble length $H^{-1}$ and the wall width $\sim m_0^{-1}$. As the universe expands, the former scales as $\tau^{2}$, while the latter stays constant. 
To ensure the validity of our results, the simulation box should contain more than ${\cal O}(1)$  Hubble horizons,
\begin{equation}
\left( \frac{a(\tau)L_{\text{box}}}{H(\tau)^{-1}} \right)^2 = \left( \frac{L_{\text{box}}}{\tau} \right)^2 \;\gtrsim\; {\cal O}(1),
\end{equation}
throughout the simulation.
Simultaneously, 
the lattice spacing given by 
$a(\tau)L_{\text{box}}/N_{\text{grid}}$ 
must be finer than the physical wall width $\sim m_0^{-1}$, 
\begin{equation}
\frac{m_0^{-1}}{a(\tau)L_{\text{box}}/N_{\text{grid}}} = \frac{N_{\text{grid}}}{m_0^2 L_{\text{box}}}\frac{1}{\tau} \;\gg \; {\cal O}(1),
\end{equation}
to prevent the momentum of the domain walls from being artificially truncated. 
Note that the actual width of the moving walls is thinner due to the Lorentz contraction. Thus, we must use sufficiently fine grid spacing to resolve the wall width, especially in the case of inflationary fluctuations.

\begin{table}[!t]
    \centering
    \begin{tabular}{>{\centering\arraybackslash}p{1.1cm}>{\centering\arraybackslash}p{1.0cm}c>{\centering\arraybackslash}p{1.4cm}cc|>{\centering\arraybackslash}p{1.1cm}>{\centering\arraybackslash}p{1.0cm}c>{\centering\arraybackslash}p{1.4cm}cc}
        \hline\hline
        Case & Fluc. & $\epsilon$ & ${N_{\rm grid}}^2$ & $m_0\tau_{\rm end}$ & Realization & Case & Fluc. & $\epsilon$ & ${N_{\rm grid}}^2$ & $m_0\tau_{\rm end}$ & Realization \\
        \hline
        (a-1)  & WN & 0      & $8192^2$  & 15 & 100     & (b-1)  & SI & 0      & $8192^2$  & 15 & 100 \\
        (a-2)  & WN & 0.0004 & $16384^2$ & 30 & 40      & (b-2)  & SI & 0.0008 & $16384^2$ & 30 & 40  \\
        (a-3)  & WN & 0.0008 & $16384^2$ & 30 & 10      & (b-3)  & SI & 0.001  & $16384^2$ & 30 & 40  \\
        (a-4)  & WN & 0.001  & $16384^2$ & 30 & 10      & (b-4)  & SI & 0.0015 & $16384^2$ & 30 & 10  \\
        (a-5)  & WN & 0.0015 & $16384^2$ & 30 & 10      & (b-5)  & SI & 0.002  & $16384^2$ & 30 & 10  \\
        (a-6)  & WN & 0.002  & $16384^2$ & 30 & 10      & (b-6)  & SI & 0.003  & $8192^2$  & 15 & 100 \\
        (a-7)  & WN & 0.003  & $8192^2$  & 15 & 100     & (b-7)  & SI & 0.005  & $8192^2$  & 15 & 100 \\
        (a-8)  & WN & 0.005  & $8192^2$  & 15 & 100     & (b-8)  & SI & 0.008  & $8192^2$  & 15 & 100 \\
        (a-9)  & WN & 0.008  & $8192^2$  & 15 & 100     & (b-9)  & SI & 0.01   & $8192^2$  & 15 & 100 \\
        (a-10) & WN & 0.01   & $8192^2$  & 15 & 100     & (b-10) & SI & 0.02   & $8192^2$  & 15 & 100 \\
        (a-11) & WN & 0.02   & $8192^2$  & 15 & 100     & (b-11) & SI & 0.03   & $8192^2$  & 15 & 100 \\
        (a-12) & WN & 0.03   & $8192^2$  & 15 & 100     & (b-12) & SI & 0.05   & $8192^2$  & 15 & 100 \\
        (a-13) & WN & 0.05   & $8192^2$  & 15 & 100     &        &    &        &           &    &     \\
        \hline\hline
    \end{tabular}
    \caption{Setting of parameters used in simulations. WN indicates the white noise case and SI indicates the scale-invariant case. All cases were run in the comoving lattice containing $(L_{\rm box})^2 = (64\pi/m_0)^2$ Hubble horizons, and in each case, the results are averaged over the number of realizations. All simulations were stopped at $\tau_{\rm end}$. In the cases with small biases, such as (a-2)-(a-6) and (b-2)-(b-5), longer simulation times are required due to their long lifetimes, so we use a larger $N_{\rm grid}$.
    }
    \label{tab:params}
\end{table}

\begin{figure}[t!]
    \begin{center}  
        \includegraphics[width=160mm]{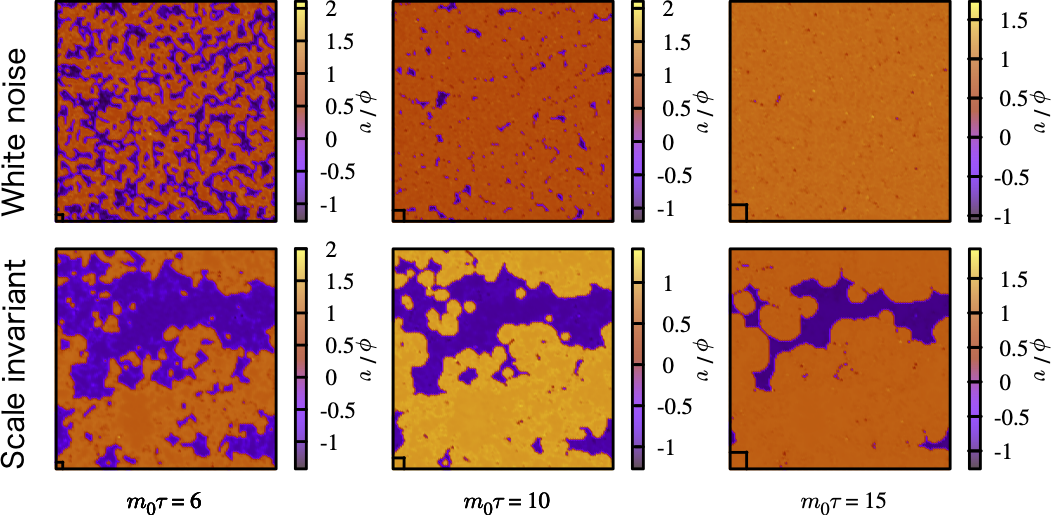}
    \end{center}
    \caption{
    Same as Fig.~\ref{fig:realspace_snapshot_unbiased} but with the potential bias, $\epsilon = 0.05$. 
    }
    
    \label{fig:realspace_snapshot_bias}
\end{figure}

\subsection{Evolution of the area parameter}
\label{subsec:evolution_of_the_area_param}

First, we show in Fig.~\ref{fig:realspace_snapshot_bias} the evolution of the domain-wall network under the influence of a potential bias. The snapshots are taken at $\tau = 6, 10$, and  $15/m_0$. With time, the pressure generated by the potential bias becomes more relevant, as the force from the tension decreases due to the cosmic expansion.
It can be seen that the closed domain walls surrounding the true vacuum regions tend to expand outward, and the system is gradually covered by true vacuum regions.

\begin{figure}[!t]
    \begin{center}  
        \includegraphics[width=160mm]{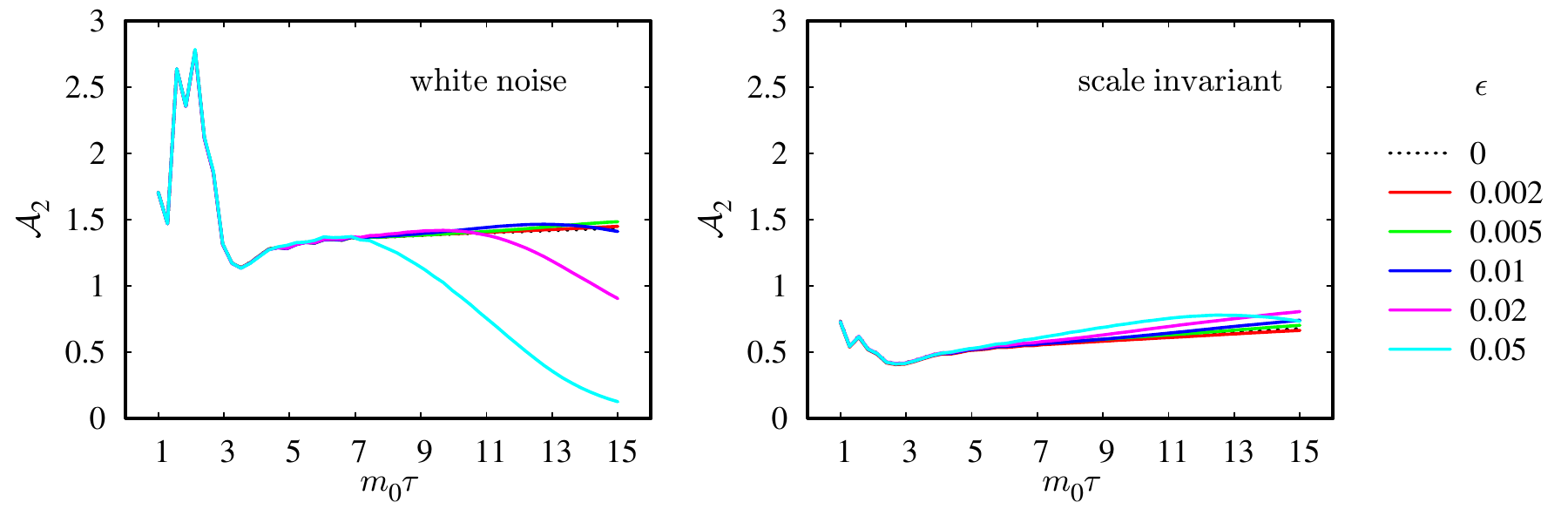}
    \end{center}
    \caption{
    The time evolution of the area parameter $\mathcal{A}_2 = L_{\rm DW}/H^{-1}$ with the bias parameter $\epsilon = 0 - 0.05$ for the white noise (left) and scale-invariant (right) initial fluctuations,
    based on 100 simulations.
    Each line represents a different bias parameter $\epsilon = 0-0.05$. The dotted lines are for the case of $\epsilon = 0$.
    }
    \label{fig:wall_length_int}
\end{figure}

Next, we examine the time evolution of $\mathcal{A}_2$, which quantifies the average domain wall area within a Hubble patch, based on our 2D lattice simulations.\footnote{In our 2D simulations, we often use common three-dimensional terminology;
the term `volume' should be understood as `area'. Similarly, when we mention the area parameter, it should be interpreted as `length'.}
We evaluate the length of walls by counting the number of lattice points on which $\left( \phi - \phi_{\rm max} \right)$ flips its sign and multiplying the lattice spacing where $\phi_{\rm max}$ is the field value for the local maximum.
In Fig.~\ref{fig:wall_length_int}, each trajectory corresponds to a different bias parameter $\epsilon$ ranging from 0 to 0.05, where the dashed line represents $\epsilon = 0$ and the solid lines represent cases with non-zero bias.

In the case of $\epsilon = 0$, the domain wall area follows the scaling law expressed in Eq.~(\ref{eq:ScalingSol2}) for $\tau \gtrsim 4/m_0$, except for a slight increase due to the finite resolution of the lattice~\cite{Gonzalez:2022mcx}.\footnote{This was further confirmed in Ref.~\cite{Gonzalez}, where it was shown that the area parameter converges to a constant as both the spatial resolution and the box size are increased. See also our Figs.~\ref{fig:wall_area} and \ref{fig:wall_area_finelat} for the results of the 3D lattice simulations and Appendix.~\ref{app:2} for 2D simulations performed with various spatial resolutions and box sizes.}
In the case of the scale-invariant fluctuations,
this increase is more pronounced when potential biases are present. We believe that this increase is mainly due to the outward expansion of the domain walls surrounding the true vacuum regions. Note that this should work only when there is plenty of space for the true vacuum regions to expand, which explains why we do not observe a pronounced increase in the white noise case. That is to say, in the white noise case, such expanding true vacuum regions soon coalesce, reducing the area parameter. On the other hand, in the scale-invariant case, there are always large voids due to the large-scale correlations, where the true vacuum regions continue to expand, increasing the area parameter. 
In this sense, the area parameter $\mathcal{A}_2$ is not a good measure of the lifetime of the domain-wall network, especially for the scale-invariant case, although it has been the focus of most of the previous literature. This is one of the reasons why we use the volume fraction of the false vacuum to study the lifetime, which will be explained next.

\subsection{Volume fraction of the false vacuum}

\subsubsection{Evolution with potential bias}

Let us introduce a volume fraction of the false vacuum, $r_V$, defined by
\begin{equation}
    r_V \;\equiv\; \frac{V_-}{V_+ + V_-},
\end{equation}
where $V_+~(V_-)$ is the volume of regions with positive (negative) values of $\phi-\phi_{\rm max}$ since the local maximum is slightly shifted by the linear potential bias. Note that in our setup the negative vacuum becomes the false vacuum in the presence of the potential bias.
The volume fraction has an advantage over the area parameter in that, averaged over realizations, it shows a monotonically decreasing function of time, which can be nicely fitted by an exponential function. 

\begin{figure}[!t]
    \begin{center}  
        \includegraphics[width=160mm]{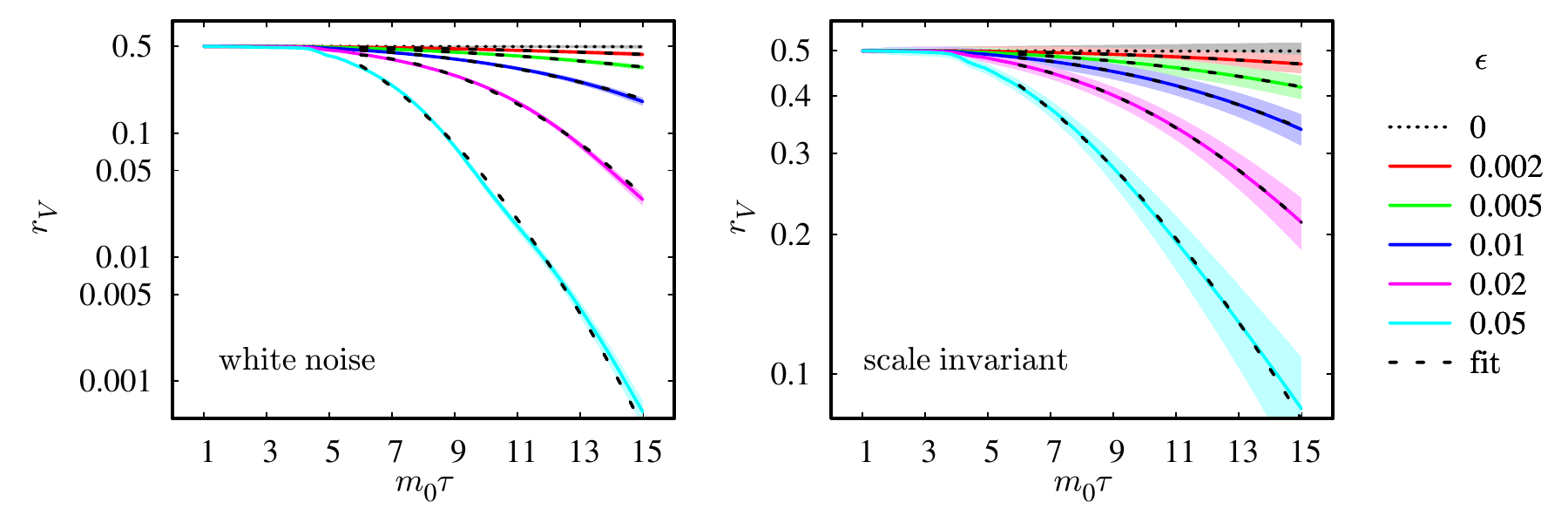}
    \end{center}
    \caption{The time evolution of the false vacuum volume fraction with the bias parameter $\epsilon = 0-0.05$ 
    for the white noise (left) and scale-invariant (right) initial fluctuations.
   The dotted lines are for $\epsilon =0$. The colored band shows the variance across realizations. 
    The dashed lines correspond to the numerical fit  assuming the
    functional form of Eq.~(\ref{eq:ansatz}). See the main text for details.
    }
    \label{fig:volume_fraction}
\end{figure}

In Fig.~\ref{fig:volume_fraction} we show the time evolution of $r_V$ for the white noise and scale-invariant initial fluctuations, varying the potential bias parameter $\epsilon$. It can be seen that, unlike the area parameter, $r_V$ decreases progressively with time for all cases, and its dependence on $\epsilon$ is also monotonic. Here, the width of the bands corresponds to the variance over realizations. The bands for the scale-invariant cases are wider than for the white noise case due to the large variance of the structure at large scales. The dashed lines are obtained from the numerical fit assuming the following dependence,\footnote{We also vary the power of $\tau-\tau'$ in the exponent, and found that the ansatz of Eq.~(\ref{eq:ansatz}) gives a good fit. 
}
\begin{equation}
    r_V \propto \exp\left[-\left(\frac{\tau-\tau'}{\tau_{\text{decay}}}\right) ^2\right].
    \label{eq:ansatz}
\end{equation}
The range for the fit is $\tau = 6-30/m_0$ for the cases (a-2)-(a-6) and (b-2)-(b-5) in Table~\ref{tab:params}, and $\tau = 6-15/m_0$ otherwise. Note that $\tau = 6/m_0$ is the time after which the scaling solution is reached when $\epsilon = 0$. The numerical fitting results are summarized in the Table~\ref{tab:DW_lifetime}.

\begin{table}[!h]
    \centering
    \begin{tabular}{c|>{\centering\arraybackslash}p{2.5cm}|>{\centering\arraybackslash}p{2.5cm}|>{\centering\arraybackslash}p{2.5cm}|>{\centering\arraybackslash}p{2.5cm}}
        & \multicolumn{2}{c|}{white noise} & \multicolumn{2}{c}{scale-invariant} \\
        \hline
        $\epsilon$ & $m_0\tau_{\rm decay}$ & $m_0\tau'$ & $m_0\tau_{\rm decay}$ & $m_0\tau'$ \\
        \hline
        0.0004 & 88.11 $\pm$ 0.55 & 1.23  $\pm$ 0.17 &                  &                  \\
        0.0008 & 54.68 $\pm$ 0.18 & -0.35 $\pm$ 0.10 & 75.84 $\pm$ 0.36 & 2.53  $\pm$ 0.13 \\
        0.001  & 50.81 $\pm$ 0.26 & 2.82  $\pm$ 0.13 & 62.33 $\pm$ 0.24 & 3.40  $\pm$ 0.10 \\
        0.0015 & 37.96 $\pm$ 0.26 & 2.36  $\pm$ 0.17 & 50.34 $\pm$ 0.15 & 3.66  $\pm$ 0.08 \\
        0.002  & 30.69 $\pm$ 0.34 & 3.38  $\pm$ 0.24 & 41.18 $\pm$ 0.35 & 3.67  $\pm$ 0.22 \\
        0.003  & 28.66 $\pm$ 0.10 & 1.52  $\pm$ 0.06 & 39.58 $\pm$ 0.16 & 2.44  $\pm$ 0.06 \\
        0.005  & 20.48 $\pm$ 0.18 & 2.37  $\pm$ 0.12 & 29.53 $\pm$ 0.16 & 2.65  $\pm$ 0.08 \\
        0.008  & 13.73 $\pm$ 0.17 & 3.59  $\pm$ 0.14 & 21.67 $\pm$ 0.14 & 3.16  $\pm$ 0.09 \\
        0.01   & 11.31 $\pm$ 0.15 & 4.10  $\pm$ 0.14 & 18.98 $\pm$ 0.14 & 3.48  $\pm$ 0.09 \\ \cline{2-3}
        0.02   & 6.10  $\pm$ 0.08 & 5.11  $\pm$ 0.10 & 12.74 $\pm$ 0.04 & 3.43  $\pm$ 0.04 \\ \cline{4-5}
        0.03   & 4.97  $\pm$ 0.05 & 4.50  $\pm$ 0.09 & 11.13 $\pm$ 0.06 & 2.49  $\pm$ 0.07 \\
        0.05   & 4.81  $\pm$ 0.13 & 1.89  $\pm$ 0.38 & 11.17 $\pm$ 0.19 & -1.02 $\pm$ 0.35 \\
    \end{tabular}
    \caption{The summary of the fitting parameters $\tau_{\rm decay}$ and $\tau'$ 
    in Eq.~(\ref{eq:ansatz}) for the 
    white noise and scale-invariant fluctuations, 
    and various potential biases. Some of the corresponding fit curves are depicted in Fig.~\ref{fig:volume_fraction}.
    Note that the errors are the asymptotic standard errors for fitting, not the variance across the realizations.
    The horizontal lines separate the strongly biased cases in which the lattice cannot resolve the domain walls well because of the Lorentz contraction.
    See the main text.
    }
    \label{tab:DW_lifetime}
\end{table}

In Table~\ref{tab:DW_lifetime}, a notable transition in behavior is observed at large values of $\epsilon$. This transition can be attributed to the limitations of the lattice resolution. In fact, as $\epsilon$ increases, the false vacuum region shrinks more rapidly, driving the domain walls to relativistic velocities earlier. In cases (a-11) to (a-13), as well as (b-11) and (b-12), an increase in the wall width at late times was observed despite the acceleration due to the large potential bias. We consider this because the lattice spacing is insufficient to resolve the Lorentz-contracted wall width.\footnote{In our numerical simulations, the (normalized) wall width is estimated as
 $n_{\rm spread}/n_{\rm flip}$,
where $n_{\rm spread}$ is a number of grid points whose field value is in the range of $ \left| \phi - \phi_{\rm max} \right| < v\tanh{\frac{1}{2\sqrt{2}}} \simeq 0.34 v$ 
and $n_{\rm flip}$ is the number of cells where the field value {$\left( \phi - \phi_{\rm max} \right)$} flips its sign.}

\begin{figure}[!t]
    \begin{center}
        \includegraphics[width=160mm]{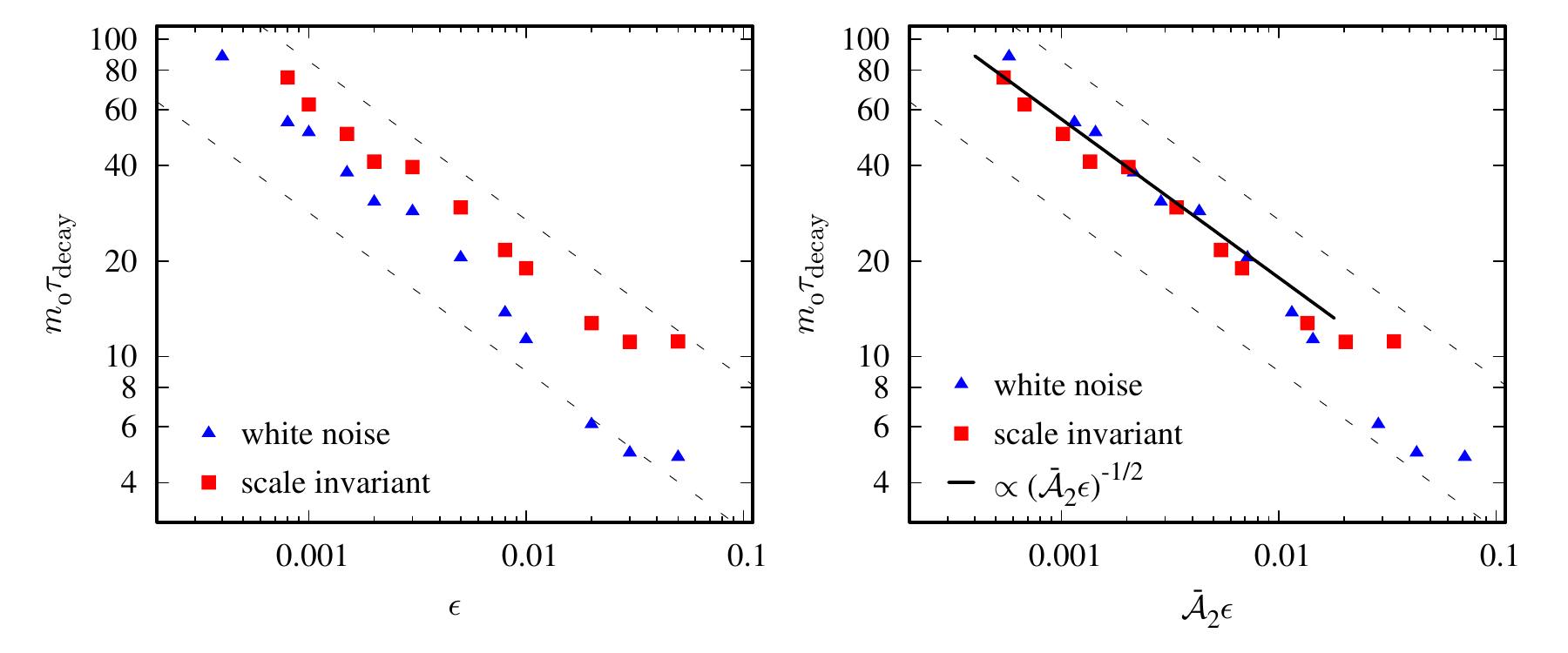}
    \end{center}
    \caption{The dependence of the false vacuum lifetime on the potential bias parameter $\epsilon$ (left) and $\bar{\mathcal{A}}_2\epsilon$ (right).
    The blue triangles are for  the white noise fluctuations and the red squares are for the scale-invariant fluctuations. The gray dashed lines $\propto \epsilon^{-1/2}$ are shown for the guidance of the eyes. In the right panel,
the black solid line is the numerical fit to the combined results with $\epsilon\leq 0.01(0.02)$ for the white-noise (scale-invariant) fluctuations.
    }
    \label{fig:A2epsilon-tau}
\end{figure}

In Fig.~\ref{fig:A2epsilon-tau} we also plot the results from Table~\ref{tab:DW_lifetime}. 
It is clear from the left panel of the figure that, given a bias, the lifetime for the scale-invariant fluctuations is longer than for the white noise fluctuations. 
We can also see from the left panel that the lifetime $\tau_{\rm decay}$ scales roughly as $\epsilon^{-1/2}$ for both types of initial fluctuations, except for large values of $\epsilon$ as mentioned above. This dependence on the bias can be understood as follows. The domain walls start to move towards the false vacuum when the pressure due to the bias becomes comparable to the tension force;
\begin{equation}
    \sigma H_{\text{eq}} = p,
    \label{eq:decay_time}
\end{equation}
where the subscript `eq' implies that it is evaluated when the above equality is met. 
During the radiation-dominated era, this leads to
\begin{equation}
    (m_0 \tau_{\text{decay}})^2 \sim 
     (m_0 \tau_{\text{eq}})^2 = 
    \frac{m_0 \tau_i}{\epsilon},
    \label{roughlifetime}
\end{equation}
and thus $\tau_{\rm decay} \propto \epsilon^{-1/2}$. This is the standard estimate of the lifetime of domain walls. However, we note that 
the condition (\ref{eq:decay_time}) determines the time when the domain wall starts to decay, and it takes some more time for the walls to disappear completely~\cite{Kawasaki:2014sqa,Saikawa:2017hiv,Murai:2023xjn}. 
In particular, note that we have not included the dependence on the area parameter in the above estimate, especially the tension force used in the definition of $H_{\rm eq}$ or $\tau_{\rm eq}$. This is partly because, as we will see shortly, the dependence of the tension force on the area parameter does not correctly lead to that of the domain wall lifetime, contrary to the estimates in Refs.~\cite{Jeong:2013oza,Kawasaki:2014sqa,Saikawa:2017hiv}. In fact, if the lifetime had the same area parameter dependence of $\tau_{\rm eq}$, the lifetime would be shorter for the inflationary fluctuations than the white noise, which is opposite to our result.

The difference between the two different initial conditions can be interpreted as arising from the global structure of the domain-wall network. To see this, in the right panel of Fig.~\ref{fig:A2epsilon-tau} we show the dependence of the lifetime on $\bar{\mathcal{A}}_{2} \epsilon$, where $\bar{\mathcal{A}}_{2}$ represents the area parameter in the absence of the potential bias, and the value of $\bar{\mathcal{A}}_2$ for the white noise and scale-invariant initial conditions is approximately 1.4 and 0.7 at $\tau = \tau_{\rm end}$, respectively.
It can be seen that the numerical results for the two cases more or less agree when plotted as a function of $\bar{\mathcal{A}}_{2} \epsilon$, except for very large values of $\epsilon$. Using the following ansatz,
\begin{equation}
    m_0\tau_{\rm decay} = \frac{\alpha}{ \sqrt{\bar{\mathcal{A}}_2 \epsilon}} ,
    \label{eq:A2epsilon-tau}
\end{equation}
we find that the numerical fit to the combined data with weak bias shown above the horizontal line in Table~\ref{tab:DW_lifetime} ($\epsilon \leq 0.01(0.02)$ for the white-noise (scale-invariant) fluctuations) leads to $\alpha = 1.77 \pm 0.05$.
For convenience, let us summarize our numerical results on the lifetime in terms of the cosmic time instead of the conformal time as follows:
\begin{equation}
\label{eq:tlife}
    t_{\rm decay} \simeq t_{\rm eq} \times 
\begin{cases}
    3 \,\sim\,5     & {\rm ~~for~white~noise}\\
    7 \,\sim\,10    & {\rm ~~for~scale~invariant}
\end{cases}
\end{equation}
where the pressure $p$ equal $\sigma H$ at $t = t_{\rm eq}$.

Note that this dependence of the domain wall lifetime on the area parameter is different from that in Refs.\cite{Jeong:2013oza,Kawasaki:2014sqa,Saikawa:2017hiv}. The references suggest a relationship of $\tau_{\rm decay} \propto \sqrt{\bar{\cal A}_2/\epsilon}$ based on an argument using  Eqs.~(\ref{eq:decay_time}) and (\ref{roughlifetime}) but with the area parameter dependence.   In other words, the area parameter was inserted into the LHS of Eq.~(\ref{eq:decay_time}). While this definition of the equality time itself is reasonable, it misses the area parameter dependence in the first equality of Eq.~(\ref{roughlifetime}). As mentioned above, the formula  in  Refs.~\cite{Jeong:2013oza,Kawasaki:2014sqa,Saikawa:2017hiv}  implies that the domain wall lifetime in the white noise case would be longer than in the scale-invariant case, which contradicts our numerical results. Therefore, $t_{\rm eq}$ is not an appropriate measure of $t_{\rm decay}$ in a qualitative sense. 

The reason why the combination $\bar{\mathcal{A}}_{2} \epsilon$ appears can be understood as follows. 
To describe the time evolution of $r_V$, we consider a simple model of the network structure. Here we consider a two-dimensional model, but the same argument can be made in three dimensions. In the following, all lengths and areas are defined in comoving coordinates, and conformal time is used. Let $\ell_{\rm DW}^{(\rm total)}$ be the ensemble average of the total length of the domain walls in the box of size $L_{\rm box}^2$. They are related to the area parameter ${\cal A}_2$ as
\begin{equation}
    \frac{\ell_{\rm DW}^{(\rm total)}}{L_{\rm box}^2} =  {\cal H} {\cal A}_2,
    \label{eq:DWlength}
\end{equation}
where we recall that ${\cal H}$ is the conformal Hubble parameter.
The change in the volume fraction $dr_V$ during the time interval $d \tau$ is caused by the motion of the domain walls, and it is estimated by
\begin{equation}
    dr_V \simeq - \frac{\ell_{\rm DW}^{(\rm total)}\times v d\tau}{L_{\rm box}^2} = -{\cal H}  {\cal A}_2  v d \tau,
    \label{eq:rVvariation}
\end{equation}
where $v$ is the averaged velocity of the domain wall along its normal direction toward the false vacuum, and we assume $v > 0$ {and $\epsilon >0$}.\footnote{The motion due to the tension force is locally random, so this evolution equation of $r_V$ should  hold even before the bias pressure dominates over the tension force.
This is the reason why $r_V$ evolves monotonically with time (see Fig.~\ref{fig:volume_fraction}), and the equality time does not directly  determine the lifetime. 
}
In the non-relativistic regime, the velocity evolves as~\cite{Avelino:2008qy}
\begin{equation}
    \frac{dv}{d\tau} \simeq \frac{p}{\sigma},
\end{equation}
where $p\, (\propto \epsilon)$ is given in Eq.~(\ref{eq:pressure}), and the cosmic expansion as well as the tension force can be neglected when $p \gtrsim \sigma H$ and the Lorentz gamma factor $\gamma = {\cal O}(1)$.  
Thus, the velocity is approximately given by $v \simeq (p/\sigma)(\tau - \tau')$, where $\tau'$ is an integration constant. Therefore,  the decrease of $r_V$ depends on a combination of ${\cal A}_2 \epsilon$. Except for very large values of $\epsilon$, it is reasonable to assume that ${\cal A}_2$ is proportional to $\bar{{\cal A}}_2$. This naive argument explains why the agreement becomes better when the lifetime is plotted as a function of $\bar{{\cal A}}_2 \epsilon$ rather than $\epsilon$.

This finding that the domain wall lifetime depends on the area parameter is new and interesting. Intuitively, the domain wall lifetime becomes longer for the inflationary fluctuations, because there are large voids and the domain walls have to traverse a larger area before they collide and annihilate. 

\subsubsection{Evolution with population bias}

For comparison, we study the time evolution of the false-vacuum volume fraction $r_V$ under the population bias.
To handle the population bias, we interpret $V_+~(V_-)$ in the definition of $r_V$ as the volume of regions with more (less) initially populated volume.
Following Ref.~\cite{Gonzalez:2022mcx}, we define the population bias parameter $b_d$ as
\begin{equation}
    b_d \equiv \left\{
    \begin{aligned}
        &\frac{\left\langle \phi \right\rangle}{\sqrt{\left\langle (\phi -\left\langle \phi \right\rangle )^2 \right\rangle}}& & {\rm ~~for \ white \ noise \ fluctuations} \\
        &\frac{\left\langle \phi \right\rangle}{\sqrt{\mathcal{P}(k)}}& & {\rm ~~ for \ scale \ invariant \ fluctuations}
    \end{aligned}
    \right. .\label{eq:popbias}
\end{equation}
Note that the variance in the denominator differs. This is to avoid the logarithmic growth of the variance in real space as a function of the box size in the case of the scale-invariant fluctuations.
We also note that the relation between the initial $r_V$ and $b_d$ depends on the details of the distribution of $\phi$. As written in Sec.~\ref{subsec4A}, our simulations choose Gaussian distributions so that both are related to the error function.

\begin{figure}[!t]
    \begin{center}  
     \includegraphics[width=160mm]{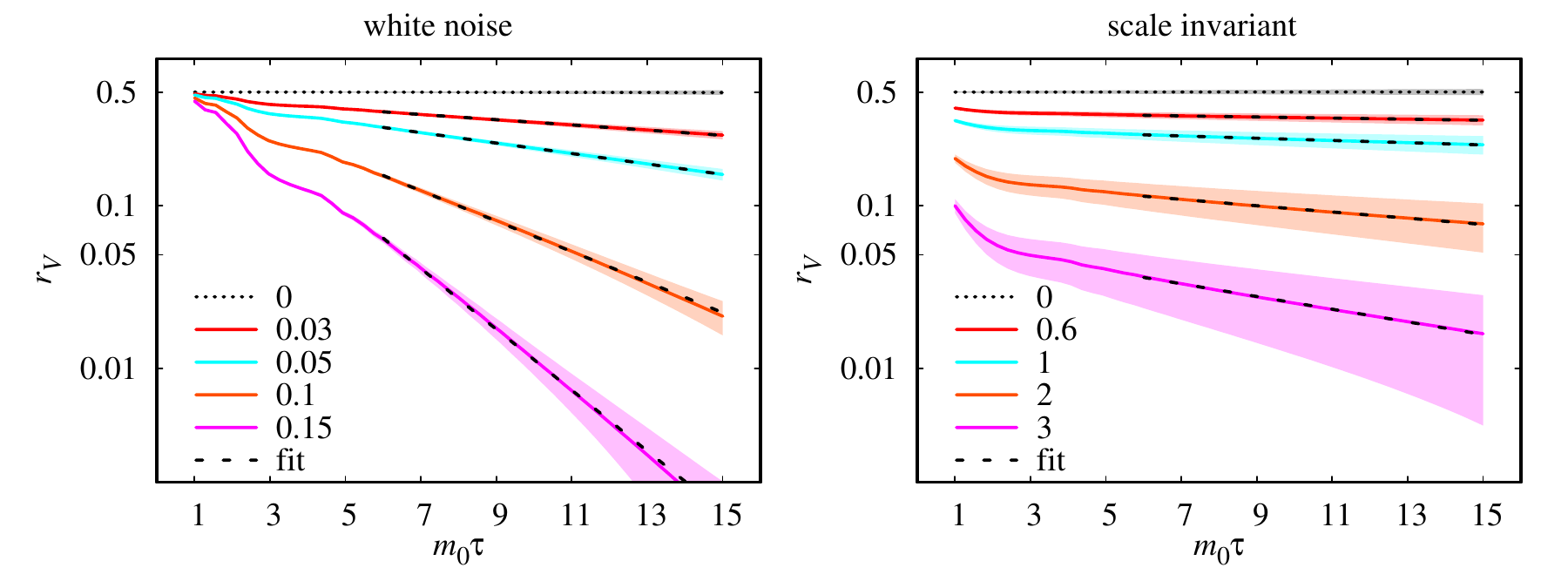}
    \end{center}
    \caption{Same as Fig.~\ref{fig:volume_fraction} but with the population bias, $b_d = 0-0.15$ for the white noise case (left) and $b_d = 0-3$ for the scale-invariant case (right).
    }
    \label{fig:volume_fraction_bd}
\end{figure}

The numerical results of the time evolution of $r_V$ are shown in Fig.~\ref{fig:volume_fraction_bd}. One can see that the domain walls are quite stable against the population bias in the case of the scale-invariant fluctuations~\cite{Gonzalez:2022mcx}. Note that the values of $b_d$ used for the two initial conditions are slightly different.
For the numerical fit, we have used the empirical ansatz,
\begin{equation}
    r_V \propto \exp\left[-\frac{\tau}{\tau_{\rm decay}} \right] \quad 
    \label{eq:ansatz_bd}
\end{equation}
which turns out to be a good fit for the numerical results. The fitting results are summarized in Table~\ref{tab:DW_lifetime_bd}.
We find that  the numerical results suggest the relation $\tau_{\rm decay} \propto {b_d}^{-1.5}$.

\begin{table}[!h]
    \centering
    \begin{tabular}{>{\centering\arraybackslash}p{1cm}|>{\centering\arraybackslash}p{2.5cm}|>{\centering\arraybackslash}p{1cm}|>{\centering\arraybackslash}p{2.5cm}}
        \multicolumn{2}{c|}{white noise} & \multicolumn{2}{c}{scale-invariant} \\
        \hline
        $b_d$ & $m_0\tau_{\rm decay}$ & $b_d$ & $m_0\tau_{\rm decay}$ \\
        \hline
        0.03 & 27.03 $\pm$ 0.08 & 0.6 & 128.41 $\pm$ 0.98 \\
        0.05 & 13.54 $\pm$ 0.01 & 1.0 & 62.56  $\pm$ 0.41 \\
        0.1  & 4.64  $\pm$ 0.02 & 2.0 & 22.53  $\pm$ 0.14 \\
        0.15 & 2.32  $\pm$ 0.01 & 3.0 & 11.13  $\pm$ 0.05 \\
    \end{tabular}
    \caption{Summary of $\tau_{\rm decay}$ defined in Eq.~(\ref{eq:ansatz_bd}) for the domain-wall network with population bias.
    }
    \label{tab:DW_lifetime_bd}
\end{table}

We have also numerically estimated the average distance between domain walls across the less populated vacuum for the two initial conditions. At $\tau = 6/m_0$ it is $\sim 0.9 H^{-1}$ for the white noise fluctuations with $b_d = 0.05$, 
while it is $\sim 2.0 H^{-1}$ for the scale-invariant fluctuations with $b_d = 1$. The difference is due to the presence of large voids in the latter case. See also Fig.~\ref{fig:realspace_snapshot_unbiased}. The structure at super-horizon scales persists throughout the evolution of the domain-wall network, which explains the longevity of domain walls with inflationary initial fluctuations against the population bias.

\subsection{Power spectrum}

The power spectrum of the scalar field value and the energy density characterize the global structure of the domain-wall network. The reduced power spectrum of the energy density ${\cal P}_\rho(k)$ is defined as {${\cal P}(k)$ for $\phi$ (see Eq.~(\ref{eq:psphi})) and it satisfies}
\begin{align}
\langle \rho({\boldsymbol x})^2 \rangle = \int d \ln k~ {\cal P}_\rho(k),
\end{align}
where the energy density $\rho$ is given by
\begin{align}
    \rho({\boldsymbol x}) = \frac{1}{2} \dot{\phi}^2 + \frac{1}{2 a^2} (\nabla \phi)^2 + V(\phi).
\end{align}
While ${\cal P}(k)$ is important for the cosmic birefringence if $\phi$ is coupled to the Chern-Simons term of photons, ${\cal P}_\rho(k)$ is relevant for the curvature perturbation and the gravitational waves.

\begin{figure}[h!]
    \begin{center}  
        \includegraphics[width=80mm]{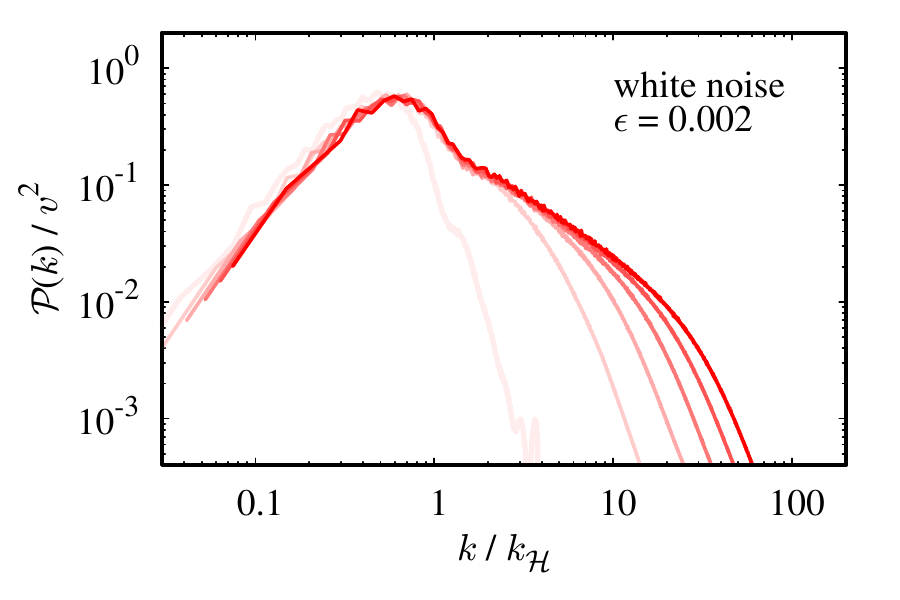} 
        \includegraphics[width=80mm]{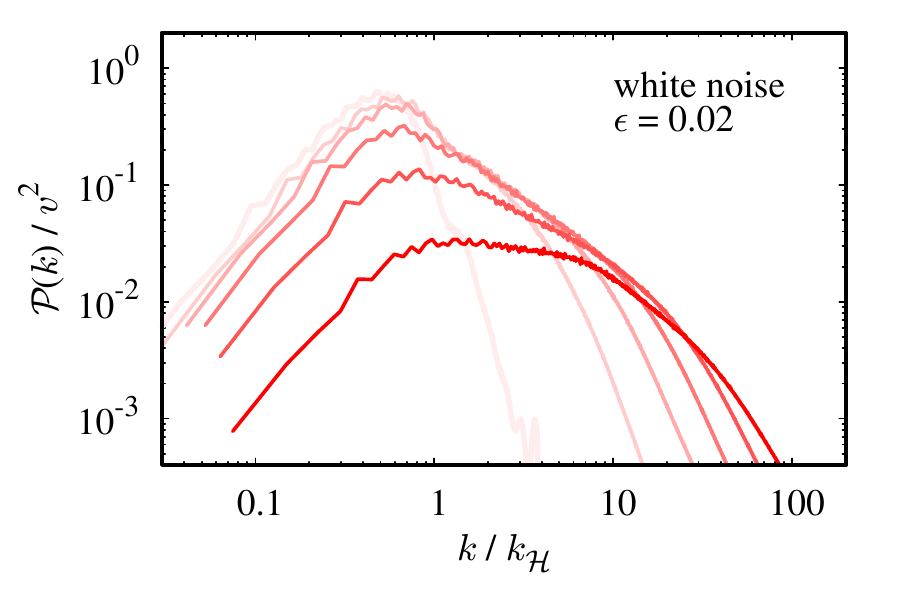}
        \includegraphics[width=80mm]{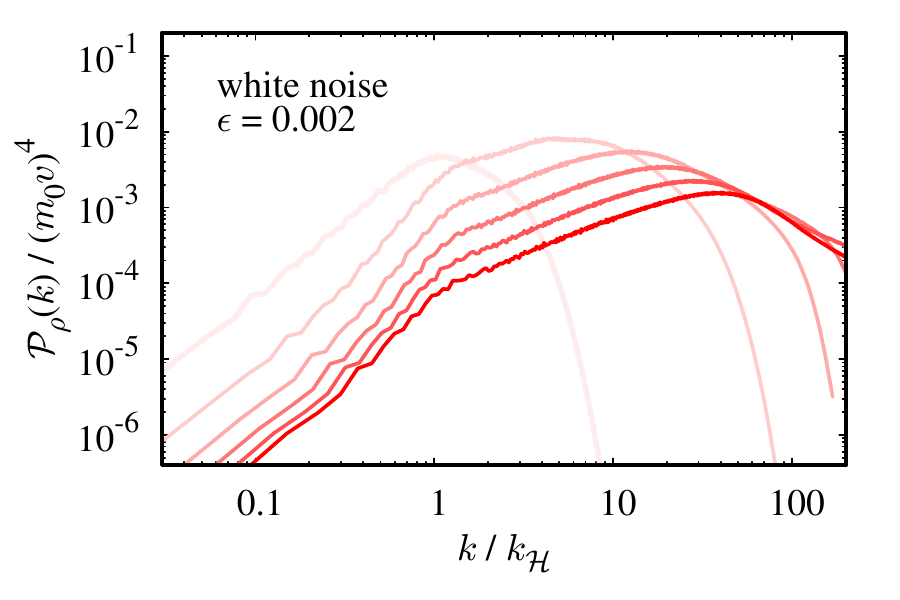}
        \includegraphics[width=80mm]{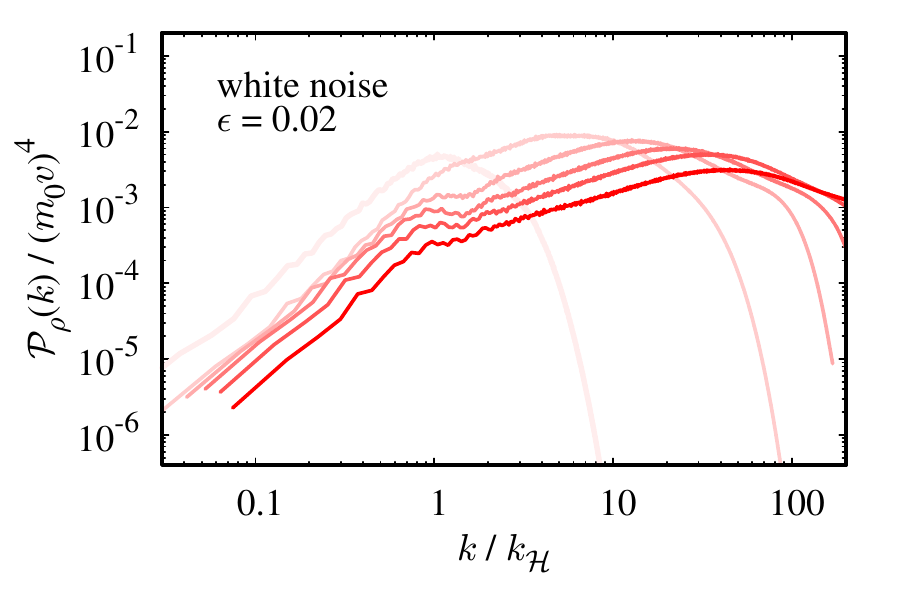}
    \end{center}
    \caption{The time evolution of the power spectrum of the field value (top) and energy density (bottom) for the white noise case with $\epsilon = 0.002$ (left) and $0.02$ (right).
    {The color of the lines indicates the temporal evolution of the results, with darker colors representing later times (equally spaced between $\tau = 4 - 15/m_0$).}
     }
    \label{fig:power_spectrum_wn}
\end{figure}

\begin{figure}[h!]
    \begin{center}  
        \includegraphics[width=80mm]{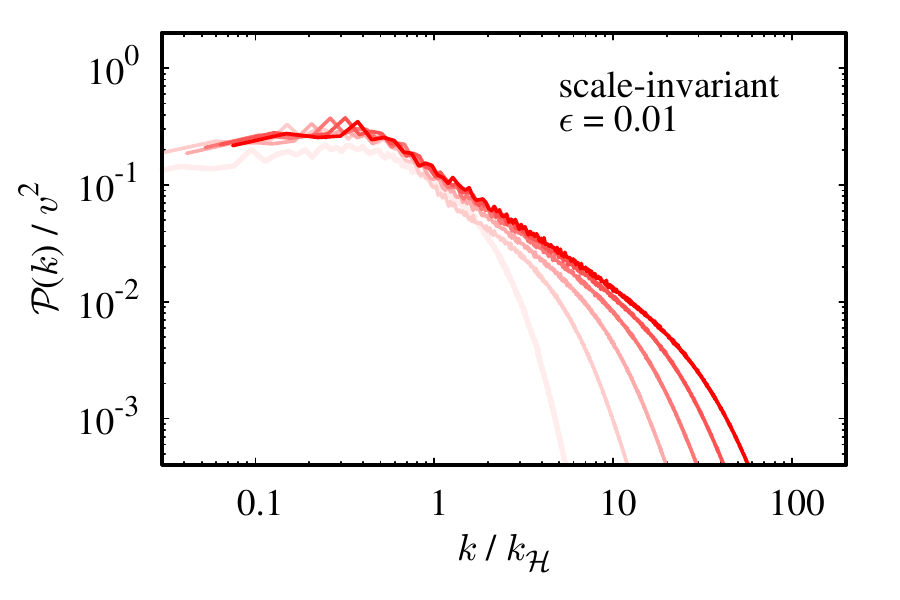} 
        \includegraphics[width=80mm]{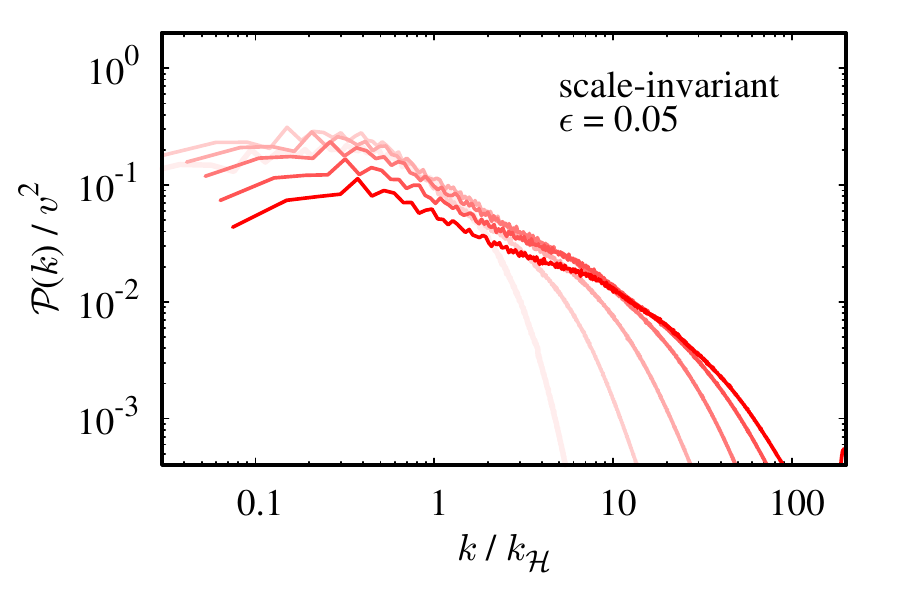}
        \includegraphics[width=80mm]{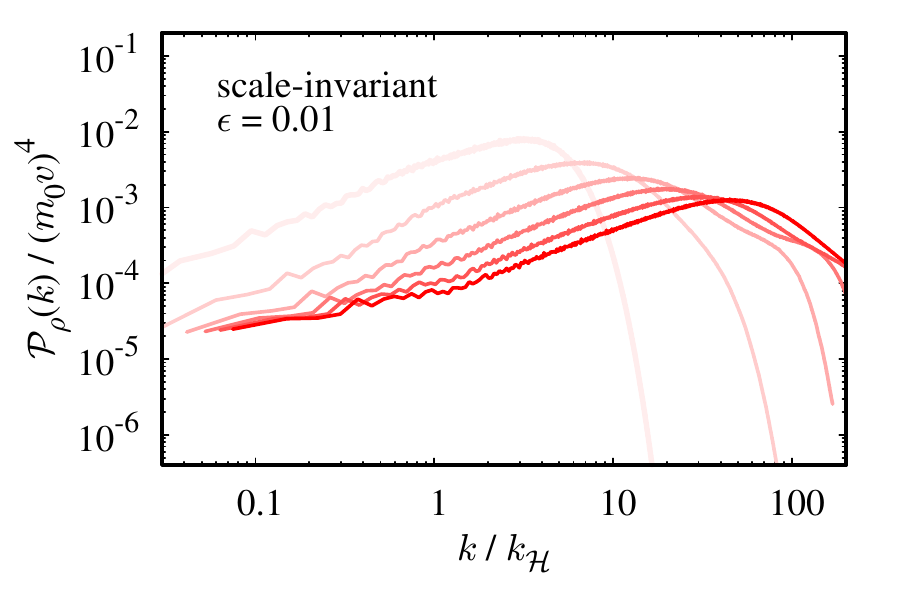}
        \includegraphics[width=80mm]{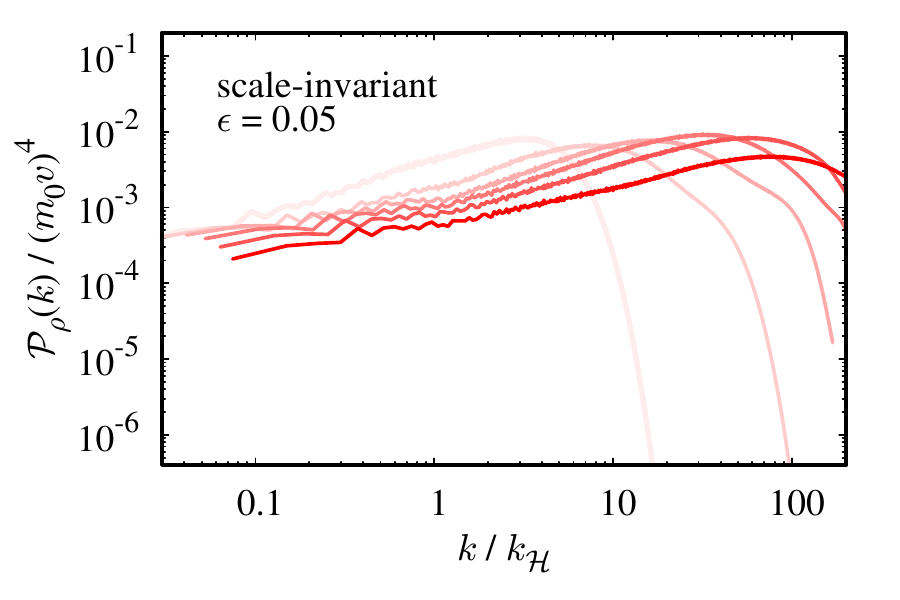}
    \end{center}
    \caption{Same as Fig.~\ref{fig:power_spectrum_wn} but for the scale-invariant case with $\epsilon = 0.01$ (left) and $0.05$ (right).
    }
    \label{fig:power_spectrum_si}
\end{figure}

In Fig.~\ref{fig:power_spectrum_wn} and \ref{fig:power_spectrum_si} we present the numerical results on the power spectra for the white noise and scale-invariant initial fluctuations, respectively. In the upper left panel of Fig.~\ref{fig:power_spectrum_wn} we can see that the power spectrum of the field value  ${\cal P}(k)$ peaks at the Hubble horizon scale, $k/k_{\cal H}=1$, and 
falls off sharply at scales smaller than the wall width, $k/k_{\cal H}=m_0^2\tau^2 / 2\pi$.  {Here $k_{\cal H} \equiv 2 \pi {\cal H}$ is the comoving wavenumber corresponding to the Hubble radius.}
This is in agreement with the scaling solution.  On the other hand, in the upper right panel with a relatively large potential bias, the power spectrum decreases with time, implying that the domain-wall network is decaying due to the potential bias. In the lower panels, we can see that the energy power spectrum also decays with time. In particular, we can see a slight enhancement at small scales when the potential bias is large. This is probably because the $\phi$ particles are produced by domain wall annihilation and they generate higher momentum modes by scattering.

In the case of the inflationary initial fluctuations, the power spectrum ${\cal P}(k)$  becomes similarly suppressed with time, especially when there is a large potential bias (see the upper right panel of Fig.~\ref{fig:power_spectrum_si}).
This is because the domains of true vacua dominate over the entire box volume, so the correlation function is strongly suppressed. On the other hand, ${\cal P}_\rho$ seems to approach a scale-invariant spectrum on the large scale at a late time. This is very different from the white noise case. The scale invariance of the energy density can be understood as follows.  When Eq.~(\ref{eq:decay_time}) is satisfied, the domain walls begin to annihilate due to the potential bias. Around this time, the power spectrum of the false vacuum energy is approximately proportional to ${\cal P}(k)$, since the energy density contrast due to the false vacuum energy is proportional to $\phi$. After this time, the domain-wall network mostly decays in {several Hubble times (see Eq.~(\ref{eq:tlife})). } The energy of the false vacuum is predominantly transferred to that of the $\phi$ particles during the decay of the domain-wall network. As a consequence, the resulting particle energy density at the end also has the scale-invariant power spectrum at the superhorizon scales, following the false vacuum power spectrum at the beginning. As we will discuss in the last section, the nearly scale-invariant energy density prevents $\phi$ from being the dominant dark matter, and it could also lead to probes of the scenario.


\section{Gravitational waves}
\label{sec5}
Domain walls continue to collide and annihilate in the scaling regime, producing gravitational waves (GWs). Here we present our numerical estimates of GWs from domain walls with the two different initial fluctuations.

The amount of GWs produced by domain walls has been studied in Refs.~\cite{Hiramatsu:2010yz,Kawasaki:2011vv,Hiramatsu:2013qaa,Higaki:2016jjh,Nakayama:2016gxi,Saikawa:2017hiv,Ferreira:2022zzo,Kitajima:2023cek}. In particular, the GWs produced during the collapse due to the potential bias were calculated for the first time by Murai and the present authors in Ref.~\cite{Kitajima:2023cek}, by considering a temperature-dependent bias. Before that, most of the works used the GWs estimated in Ref.~\cite{Hiramatsu:2013qaa} where they did not include the potential bias term in the numerical lattice calculations. 
The omission of the bias term was primarily due to the extended lifetime of the domain-wall network under a bias that ensures the system achieves a scaling solution well before annihilation. Additionally, accurately simulating the annihilation process within the constraints of numerical simulation time posed significant challenges.
Here our focus is on the difference between the white noise and scale-invariant initial fluctuations, so we estimate the GWs during the scaling regime without introducing the potential bias term, following the spirit of Ref.~\cite{Hiramatsu:2013qaa}. The GW spectrum after the domain wall collapse can be approximately estimated by evaluating the GW in the scaling regime at the time of the decay, determined by Eq.~(\ref{eq:decay_time}).\footnote{
Note, however, that the main production of the GWs occurs during the collapse of domain walls~\cite{Kitajima:2023cek}.
}
We leave it to future work to estimate the GWs emitted during the collapse for the inflationary initial fluctuations.

The tensor perturbation of the metric in a spatially flat FLRW universe is given by
\begin{equation}
    ds^2 = -dt^2 + a^2(t) (\delta_{ij}+h_{ij})dx^i dx^j,
\end{equation}
where 
$h_{ij}$ is the transverse and traceless tensor with $|h_{ij}| \ll 1$.
The equation of motion for $h_{ij}$ is
\begin{align}
    \ddot{h}_{ij} + 3H \dot{h}_{ij} - \frac{1}{a^2}\Delta h_{ij} = \frac{2}{M_{\rm pl}^2} {\Lambda_{ij}}^{kl}T_{kl},
\end{align}
where $M_{\rm pl}$ is the reduced Planck mass, 
$T_{ij}$ is a stress tensor, and ${\Lambda_{ij}}^{kl}$ is a projection onto the transverse and traceless components of the stress tensor. In our scenario, we use the stress tensor for the scalar field.

The GW spectrum is estimated as
\begin{align}
    \Omega_{\rm GW}(k,t) &= \frac{1}{\rho_c (t)}    \rho_{\rm GW}(k,t)    
    \label{eq:defofOmegaGW}\\
    \rho_{\rm GW}(k,t) &\equiv 
    \frac{M_{\rm pl}^2}{4}\left\langle{\dot{h}_{ij}(k,t) \dot{h}_{ij}(k,t)}\right\rangle, 
\end{align}
where 
$\Omega_{\rm GW}(k,t)$ and $\rho_{\rm GW}(k,t)$ are the density parameter and the energy density of GWs per the unit logarithmic wavenumber, respectively, $\rho_c(t) = 3H^2 M_{\rm pl}^2$ is the critical density at time $t$, and the bracket represents the spatial average. 
Let us introduce the so-called efficiency parameter $\epsilon_{\rm GW}$~\cite{Hiramatsu:2013qaa},
\begin{equation}
    \rho_{\rm GW}(k,t) = \frac{\epsilon_{\rm GW}(k,t) {\cal A}_3^2\sigma^2}{32\pi M_{\rm pl}^2},
\end{equation}
where $\epsilon_{\rm GW}$ should be of order unity around the peak frequency in the scaling regime, and ${\cal A}_3$ is the area parameter in the absence of any biases.

\begin{figure}[t!]
    \begin{center}  
        \includegraphics[width=80mm]{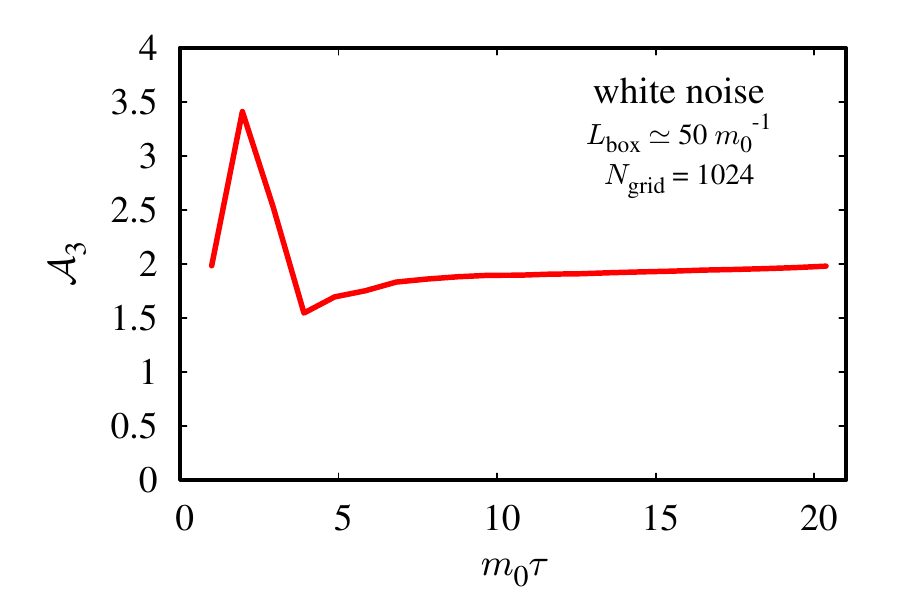}
        \includegraphics[width=80mm]{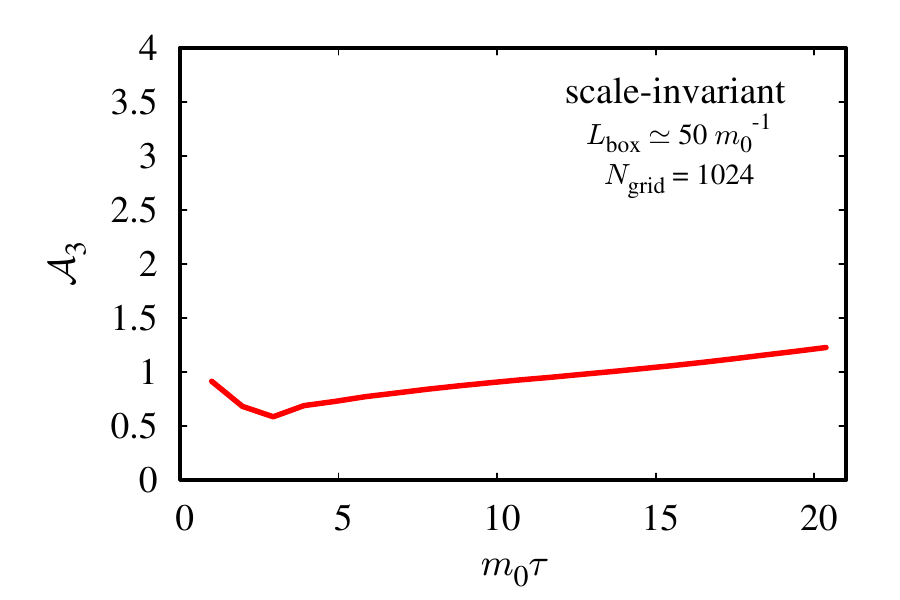}
    \end{center}
    \caption{
    The time evolution of ${\cal A}_3$ without any biases for the white noise (left) and scale-invariant (right).
    We use the lattice with $N_{\rm grid}^3 = 1024^3$ containing $\sim 50^3$ Hubble horizons at the initial time.
    }
    \label{fig:wall_area}
\end{figure}

In the scaling regime, the typical distance between domain walls is of order the Hubble radius.
The GW spectrum has a peak at the corresponding scale,
\begin{equation}
   k_{\rm peak} \sim k_{\cal H} = 2 \pi a H
    \label{eq:GWpeakfreq}
\end{equation}
with the GW density parameter,\footnote{
Note that the dependence of $\Omega_{\rm GW}$ on the area parameter is consistent with Refs.~\cite{Hiramatsu:2013qaa,Saikawa:2017hiv},
because we did not consider the GW emission during the collapse due to potential bias. The actual dependence on the area parameter may be more involved. 
}
\begin{equation}
    \Omega_{{\rm GW}}(k_{\rm peak},t) = \frac{\tilde{\epsilon}_{\rm GW}\mathcal{A}_3^2\sigma^2}{96\pi M_{\rm pl}^4 H^2},
    \label{eq:GWpeakabun}
\end{equation}
where $\tilde{\epsilon}_{\rm GW}(t) \equiv {\epsilon}_{\rm GW}(k_{\rm peak},t)$, and it was estimated to be $\tilde{\epsilon}_{\rm GW}\simeq 0.7\pm 0.4$ in Ref.~\cite{Hiramatsu:2013qaa}.
We have seen that the value of ${\cal A}_2$ differs by a factor of 2 between the white noise and scale-invariant fluctuations in the case of the two-dimensional simulations. See the line of $\epsilon = 0$ in Fig.~\ref{fig:wall_length_int}.
We have confirmed that this is also the case for the three-dimensional case. See Fig.~\ref{fig:wall_area}.  We can see that the evolution of ${\cal A}_3$ is very similar to that of ${\cal A}_2$. 
Therefore, the peak frequency for the scale-invariant fluctuations is expected to be lower by a factor of about two than for the white noise fluctuations.

\begin{figure}[t!]
    \begin{center}  
        \includegraphics[width=80mm]{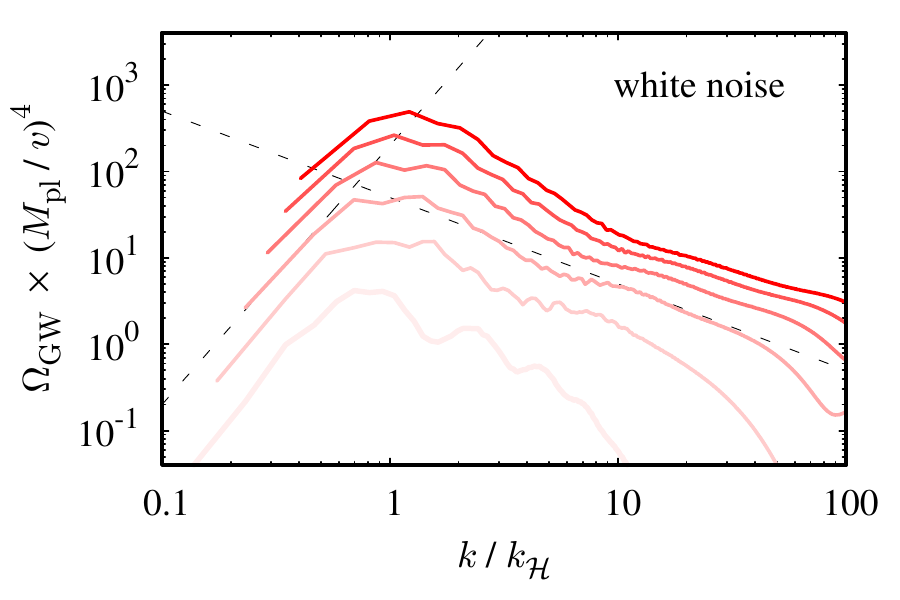}
        \includegraphics[width=80mm]{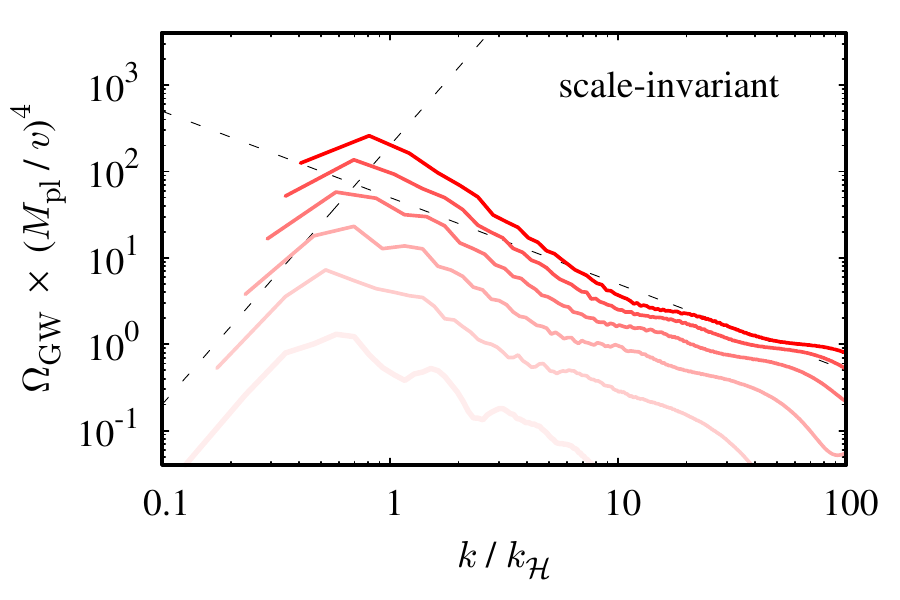}
    \end{center}
    \caption{
    GW spectra from 15 realizations for the white noise (left) and scale-invariant (right) cases, with no potential bias. The color of the lines indicates the time evolution within $\tau = 6 - 20/m_0 $, with darker colors corresponding to later times. Black dashed lines represent power laws $ \propto k^3 $ and $ \propto k^{-1} $, respectively, to guide the eye.
    }
    \label{fig:spectrumGW}
\end{figure}

We have performed numerical simulations on a three-dimensional lattice with $N_{\rm grid} = 1024$. The box contains $\sim 50^3$ Hubble horizons at the initial time $\tau_i = 1/m_0$. As we mentioned, we do not introduce potential biases, setting $\epsilon=0$. 
The simulation is stopped at a time $\tau_{\rm end}=20/m_0$.  In Fig.~\ref{fig:spectrumGW} we show the GW spectra obtained by averaging the results of 15 simulations. These simulations were performed with the white noise (left) and the scale-invariant (right) initial fluctuations. 
The color of the lines represents the time evolution, with darker colors corresponding to data at later times. Specifically, we show the data at $\tau \simeq 6, 9, 12, 15, 17,$ and $ 20/m_0$. 

\begin{figure}[t!]
    \begin{center}  
    \includegraphics[width=80mm]{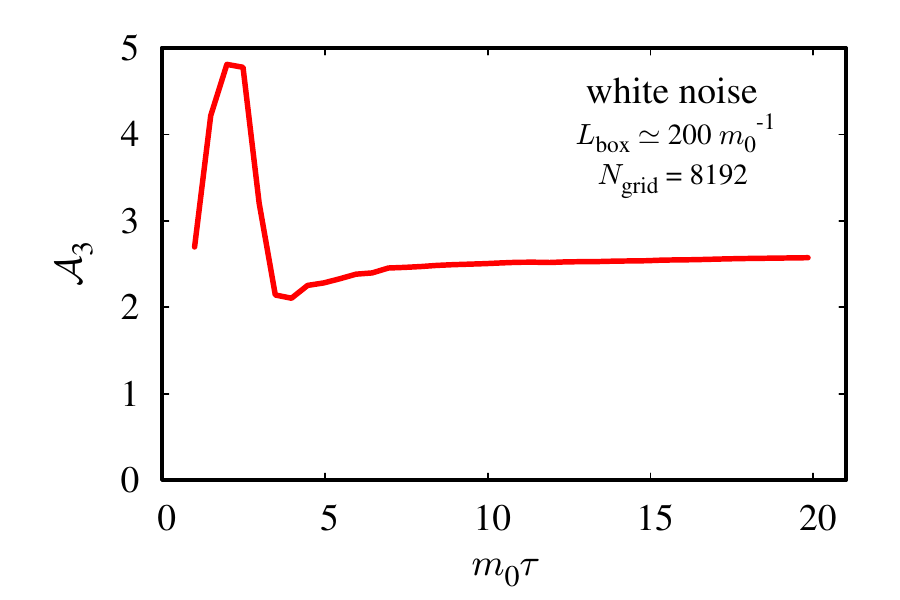}
    \includegraphics[width=80mm]{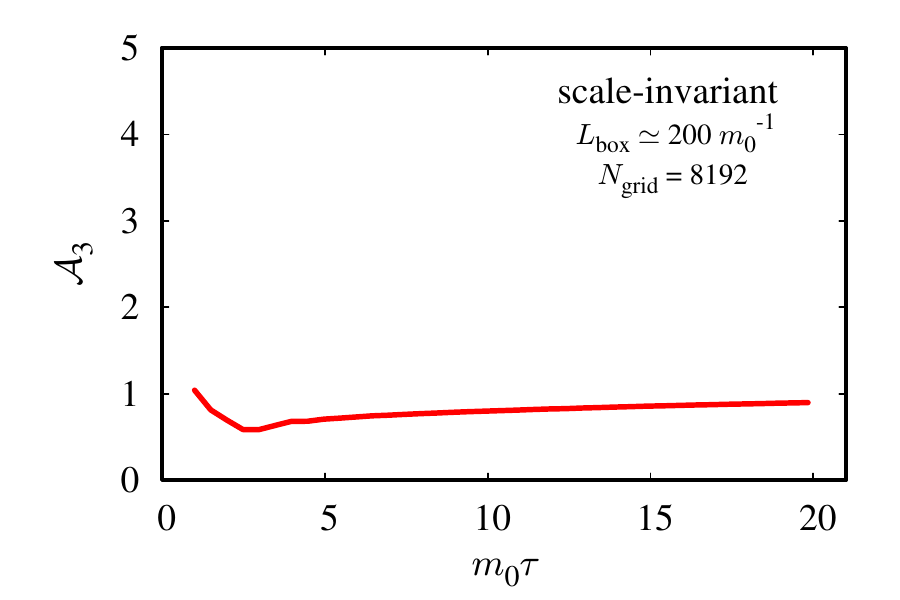}
    \end{center}
    \caption{
    Same as Fig.~\ref{fig:wall_area} but with $N_{\rm grid}^3 = 8192^3$ containing $\sim 200^3$ Hubble horizons at the initial time.
    }
    \label{fig:wall_area_finelat}
\end{figure}

Here we comment on the increase of ${\cal A}_3$ in time observed for the scale-invariant case in Fig.~\ref{fig:wall_area}. This is due to the limitation of the simulation box size (see also the discussion in~Ref.\,\cite{Gonzalez}).
More precisely, for the superhorizon scale correlations to be properly incorporated, the simulation box should contain at least several horizons.
In our case, however, the box at the final time contains only $\sim 2.5^3$ horizons, which should be the reason for the  increase of the ${\cal A}_3$.
To verify this understanding, we performed simulations with a larger number of lattice, $N_{\rm grid}^3 = 8192^3$ in Fig.~\ref{fig:wall_area_finelat}. The box contains $\sim 200^3$ horizons at the initial time and $\sim 10^3$ horizons at the final time.\footnote{This simulation has been conducted by 
supercomputer AOBA in Tohoku University.}
As shown in Fig.~\ref{fig:wall_area_finelat}, the increase in ${\cal A}_3$ for the scale-invariant case is not pronounced as the smaller grid case and follows similar behavior with white noise case after $\tau \simeq 4 / m_0$, except for the overall magnitude.
See also Appendix.~\ref{app:2} for the volume effect on the area parameter in the 2D simulations.
Therefore, we showed that the small area parameter of the domain-wall networks reflects the influence of superhorizon scale correlations rather than some artifacts from the numerical simulations.\footnote{That said, we admit that the area parameter exhibits mild time dependence for both initial conditions. To confirm that the difference in the area parameter is a genuine physical effect rather than a transient one, simulations with a much larger dynamical range
would be required, which is technically challenging at present.}


The comparison of the GW spectra for both cases shows a similarity in the overall shapes. This should be contrasted with the 
power spectra of the energy density, which are clearly different at large scales. This implies that the energy density distribution at superhorizon scales does not lead to the production of GWs, which is consistent with our intuition.

Looking more closely at the spectrum, both spectra show the same power-law behavior at large scales ($\propto k^3$) and small scales ($\propto k^{-1}$), respectively.
This wavenumber dependence agrees with the finding of Ref.~\cite{Hiramatsu:2013qaa} which adopted the initial fluctuations of a massless scalar field in the Minkowski space.
However, we can see that the peak shifts toward lower frequencies, resulting in reduced GWs for the scale-invariant case. The difference can be attributed to the difference of ${\cal A}_3$. The peak frequency corresponds to the typical correlation length of the domain-wall network, and the GW emission depends on the total energy stored in the network as well as the curvature radius of domain walls. 

\begin{figure}[t!]
    \begin{center}  
        \includegraphics[width=80mm]{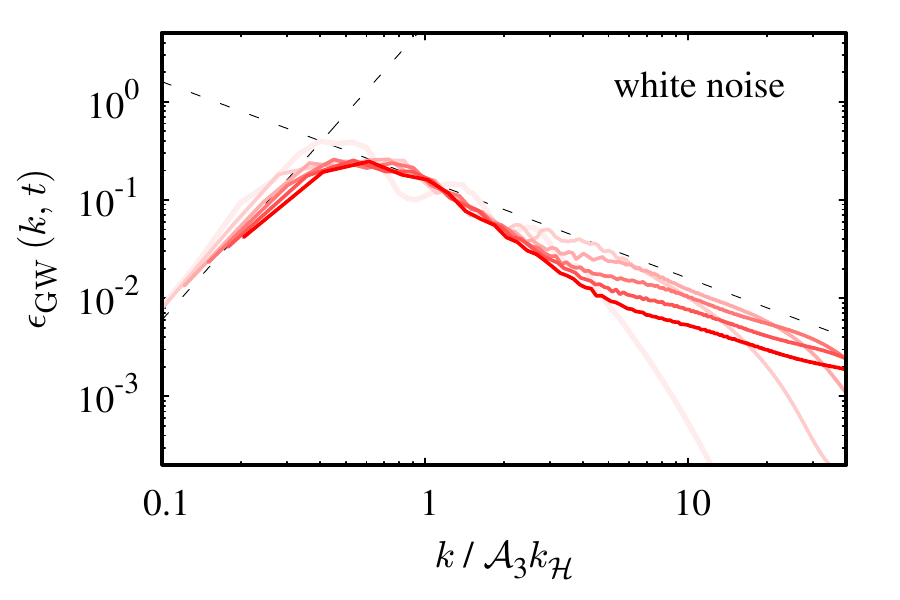}
        \includegraphics[width=80mm]{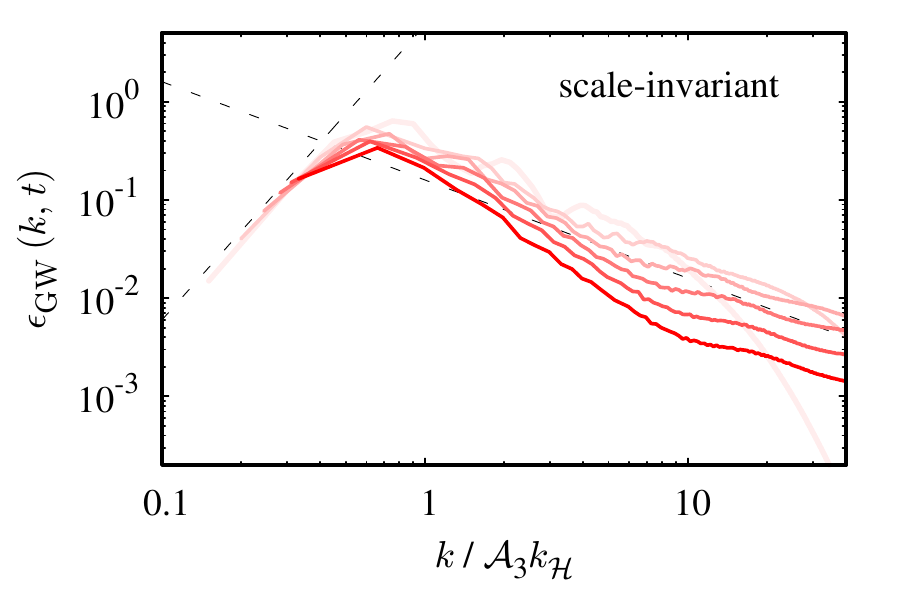}
    \end{center}
    \caption{Same as Fig.~\ref{fig:spectrumGW} but the horizontal axis is re-scaled by ${\cal A}_3(\tau)$ and the vertical axis is the efficiency parameter ${\epsilon}_{\rm GW}(k,t)$.     }
    \label{fig:spectrumGW_mod}
\end{figure}

In Fig.~\ref{fig:spectrumGW_mod}, we show the spectrum of the efficiency parameter with respect to the wavenumber rescaled by ${\cal A}_3 k_{\cal H}$, which corresponds to the typical distance scale between domain walls. Note that we take account of the time-dependence of ${\cal A}_3$. We can see that at any given time, the spectrum around the peak frequency behaves almost the same without much dependence on the two initial conditions.
Choosing the data at $\tau = \tau_{\rm end}$, our results show $\tilde{\epsilon}_{\rm GW}$ to be
\begin{equation}
    \label{eq:epsilonGWtilde}
    \tilde{\epsilon}_{\rm GW} = 
    \begin{cases}
        0.25 \pm 0.04 & {\rm ~~for~white~noise    }\\
        0.34 \pm 0.14 & {\rm ~~for~scale~invariant}
    \end{cases}
    ,
\end{equation}
where the errors come from variations of $\Omega_{\rm GW}$ and ${\cal A}_3$ across realizations.
Therefore, our results show that by considering the dependence of the peak height and peak frequency of the GW spectrum on the area parameter, we can predict the gravitational wave spectrum independently of the time and nature of the initial fluctuations.
In particular, based on our results, we improve the estimate of the peak wavenumber given by Eq.~(\ref{eq:GWpeakfreq}) as follows
\begin{equation}
    k_{\rm peak} \simeq {\cal A}_3 k_{\cal H} = 2\pi a {\cal A}_3 H ,
    \label{eq:GWpeakfreq_mod}
\end{equation}
under the scaling regime.
The peak wavenumber (\ref{eq:GWpeakfreq_mod}) takes into account the area parameter dependence, which reflects the global structure of the domain-wall network.
Note, however, that if ${\cal A}_3$ is much less than unity, e.g. due to dilution by secondary inflation, we expect this relationship to break down due to the limitation of causality.

Lastly, let us compare our results of $\tilde{\epsilon}_{\rm GW}$ with the past literature. In fact, these values in Eq.~(\ref{eq:epsilonGWtilde}) are smaller than those of Ref.~\cite{Hiramatsu:2013qaa}.
However, considering the difference in the method of evaluating the area parameter as well as the initial condition, 
it is not straightforward to compare our results with theirs. 
While we use the white noise initial fluctuations, Ref.~\cite{Hiramatsu:2013qaa} adopted fluctuations of a massless scalar field in the Minkowski space.
Apart from the different initial conditions,
our method tends to count the area of walls more than the reference. Specifically, Ref.~\cite{Hiramatsu:2013qaa} gives ${\cal A} \simeq 0.8\pm 0.1$, which would imply ${\cal A}_3 \simeq 1.6\pm 0.2$ in the non-relativistic approximation. On the other hand, we have ${\cal A}_3 \simeq 2$ for the white noise fluctuations, which makes our estimate of $\tilde{\epsilon}_{\rm GW}$ smaller by a factor of $1.6$. Given this difference in the estimate of the area parameter, our result for the white noise fluctuations is consistent with Ref.~\cite{Hiramatsu:2013qaa}, {although such a direct comparison may not be justified since different initial conditions are used.}

\section{Discussion and conclusions}
In this paper we have studied the evolution of the domain-wall network in the presence of the potential and population biases for the two different initial fluctuations, thermal and inflationary fluctuations. In particular, this is the first to study the domain-wall network with the inflationary fluctuations under the influence of the potential bias.  While in our previous work the domain-wall network was found to be very robust against the population bias for the inflationary fluctuations, we find that the domain walls decay under the potential bias with a lifetime several times longer than for the thermal fluctuations. Importantly, we have found that this difference of the lifetime arises from the area parameter\footnote{Note that the scaling behavior of domain walls characterized by the area parameter needs further numerical studies to better understand its physical nature of initial condition dependence. We assume that the deviation of the area parameter after $\tau \gtrsim 6/m_0$ is not transient but intrinsic and domain-wall network reaches the scaling regime in both cases here. However, the conclusion that large voids in the scale-invariant case contribute to the longer lifetime of domain walls does not change regardless of whether the area parameter increases with time or not.} since it determines the averaged distance between domain walls. This dependence on the area parameter has been overlooked in the literature, which has focused only on thermal initial fluctuations or similar. This led us to derive a improved formula for the domain wall lifetime which depends on the area parameter in a qualitatively different way from previous studies. One of the important lessons we have learned in this work is that the choice of initial fluctuations is important for the evolution of domain walls.

We have estimated the gravitational waves generated by domain wall annihilation for both types of initial fluctuations. The resulting gravitational wave spectrum has a peak frequency around the Hubble horizon at the time of annihilation, but the peak location is different due to the different lifetimes for the two types of initial fluctuations. We have shown that the gravitational wave spectrum can be understood in a unified manner by taking account of the dependence of the peak height and frequency of the spectrum on the area parameter. 

One important application of our results is the axion domain walls coupled to photons, which can naturally explain the isotropic cosmic birefringence~\cite{Takahashi:2020tqv} recently suggested by analyses in Refs.\cite{Minami:2020odp,Diego-Palazuelos:2022dsq,Eskilt:2022wav}. A crucial assumption is the absence of cosmic strings attached to these domain walls. If cosmic strings were present, anisotropic cosmic birefringence would likely dominate, making it challenging to account for the observed isotropic cosmic birefringence as discussed in Refs.~\cite{Takahashi:2020tqv,Jain:2021shf}. 
As discussed in the Appendix.~\ref{app:1}, such axion domain walls can form naturally from the inflationary fluctuations that the axion acquired during inflation.
While there is an inherent population bias,  the axion domain walls are stable against the population bias~\cite{Gonzalez:2022mcx}.\footnote{
Note that it is the population bias defined on the last scattering surface, $b_d^{\rm LSS}$, that is relevant for isotropic cosmic birefringence, while it is generally different from $b_d $ defined in the three-dimensional universe (see Eq.~\ref{eq:popbias}) due to the stochastic nature of the domain walls. However, for $b_d\lesssim {\cal O}(1)$, which is required for the stability of the axion domain walls, $b_d^{\rm LSS}$ is also expected to satisfy $b_d^{\rm LSS} \lesssim {\cal O}(1)$. This population bias implies an uncertainty in the predicted rotation angle, but still favors the axion-photon anomaly coefficient of ${\cal O}(1)$ to explain the observed cosmic birefringence.
}
However, they might still be subject to potential bias if the axion potential receives additional contributions. To explain the isotropic cosmic birefringence, the potential bias must be sufficiently small so that the axion domain walls exist until the time of recombination. In particular, the pressure due to the potential bias should be smaller than $\sigma H_{\rm rec}$, where $H_{\rm rec} \simeq 3 \times 10^{-29}{\rm\,eV}$ is the Hubble parameter at the time of recombination. 
This implies that the potential energy difference between the true and false vacua must be smaller than the product of the height of the axion potential and the ratio of the Hubble parameter at recombination to the axion mass.

We have primarily considered domain wall collapses under the influence of biases. After the collapse of the domain wall, most of the energy is converted into $\phi$ particles.
These particles soon become non-relativistic and may contribute to dark matter. 
In fact, we have numerically confirmed that the total energy density does indeed scale as non-relativistic matter after the domain walls have largely collapsed. Such non-thermal production of dark matter can be an important production mechanism, especially when the particles are light. Therefore, our study has important implications for the production of dark matter by domain wall collapse. However, one must pay attention to the initial fluctuations. This is because, if the domain wall originates from the inflationary fluctuations, the resulting particle energy density is expected to have a scale-invariant power spectrum at large scales~(see Fig.~\ref{fig:power_spectrum_si}). This contributes to the isocurvature perturbations with large non-Gaussianity, which would be inconsistent with observations if it were a significant fraction of dark matter. A similar argument implies that if the domain wall originates from the inflationary fluctuations, the energy density of the domain wall must be much smaller than the other components when it decays, since otherwise it would generate too large curvature perturbations with sizable non-Gaussianity.\footnote{Therefore, the GWs from axion domain walls with initial inflationary fluctuations are unable to explain the NANOGrav results~\cite{NANOGrav:2023hvm}. See also Refs.~\cite{Higaki:2016jjh,Ferreira:2022zzo,Kitajima:2023cek}.} Thus, the cosmological domain wall problem is more severe for the domain walls from the inflationary initial  fluctuations.  From another point of view, the isocurvature perturbations with significant non-Gaussianity could be a probe of the domain walls from the inflationary fluctuations. In particular, axion domain walls can be formed naturally by inflationary fluctuations in string theory, which contains many axions. Thus, the axion domain wall and its observational implications might be used to probe the string theory.

\section*{Acknowledgments}
This work is supported by  JSPS Core-to-Core Program (grant number: JPJSCCA20200002) (J.L. and F.T.), 20H01894 (F.T. and  N.K.), 20H05851 (F.T.,  N.K. and W.Y.), 21H01078 (N.K.), 21K20364 (W.Y.),  22K14029 (W.Y.), 22H01215 (W.Y.), Graduate Program on Physics for the Universe (J.L.), and Watanuki International Scholarship Foundation (J.L.). This article is based upon work from COST Action COSMIC WISPers CA21106, supported by COST (European Cooperation in Science and Technology). Part of the results in this work were obtained using supercomputing resources at Cyberscience Center, Tohoku University.

\appendix

\section{Phase transition with inflationary fluctuations}
\label{app:1}
We introduce a novel mechanism for domain wall/string formation that is similar to the Kibble mechanism~\cite{Kibble:1976sj}.  Instead of utilizing the thermal potential to establish a symmetric phase, we exploit the effects of {the stochastic dynamics}, allowing  the presence of a non-minimal coupling
 As we will see, this mechanism leads to phenomena that differ significantly from the Kibble mechanism and naturally result in the formation of topological defects through inflationary fluctuations.  A similar mechanism was mentioned in Ref.\,\cite{Hodges:1991xs} in the context of string/texture formation and in Ref.\,\cite{Nagasawa:1991zr} with a time-dependent Hubble parameter during inflation. 
However, this study examines domain-wall network formation 
{using a Hubble-induced mass with an almost constant Hubble parameter during inflation}, which has never been explored before.
Since the non-minimal coupling will be generated radiatively, it is natural to take it into account.
{This non-minimal coupling and stochastic dynamics make domain wall formation by inflationary fluctuations a very plausible scenario.}
In the case of the axion domain wall, we examine the parameter range in which strings do not form. This allows for an explanation of isotropic cosmic birefringence~\cite{Takahashi:2020tqv}.

The details of the phase transition depend on the UV completion of the models, so in the following we consider the $Z_2$ model and the axion separately.

\subsection{Case of $Z_2$ domain wall}
Here we present a simple model similar to the usual domain wall formation associated with a phase transition.  We expect analogous  domain wall formation in a larger class of models. 
We {consider} a non-minimal coupling 
\begin{equation}
{\cal L}= -c_H \frac{\phi^2R}{2}
\end{equation}
with $c_H>0$, where $R$ is the Ricci scalar given by  $R=6 (\ddot{a}/a+(\dot{a}/a)^2)$ in the flat FLRW universe. We note that a very small non-minimal coupling  is unnatural if $\phi$ has interaction because it is generated radiatively.
During inflation, the field $\phi$ acquires a positive Hubble-induced mass squared,
\begin{equation}
m_{\rm Hubble}^2= 6 c_H H_{\rm inf}^2,
\end{equation}    
where $H_{\rm inf}$ is the Hubble parameter during the inflationary period as in the main part.
With the quartic coupling or other interactions, we also have a ``thermal" mass squared of ${\cal O}(\lambda) H_{\rm inf}^2$ induced by the Gibbons-Hawking radiation. We include such additional contribution(s) by redefining $c_H$ and call the two contributions Hubble-induced mass. 
In the following, we assume $m_{\rm Hubble}^2 > m_0^2$ so that the effective potential is minimized at $\phi=0$. 
Then the system is similar to the usual Kibble mechanism where the SSB is induced by the time evolution of the thermal potential. The only difference is that the system is lifted into the symmetric vacuum by the Hubble-induced mass, but not by the real thermal mass.
Below, we consider two cases with respect to the magnitude of $c_H$.

\paragraph{$c_H\gtrsim (0.1)$}
In the case $c_H\gtrsim (0.1)$, we have the positive $m^2_{\rm Hubble}$ mass larger than {$H^2_{\rm inf}$}, and thus the inflationary fluctuation is suppressed at superhorizon scales.  
After the inflation, the Hubble parameter $H$ starts to decrease.
{When the time-dependent effective mass determined by the sum of $m_{\rm Hubble}^2$ and $V''(\phi)$  becomes comparable to $H$}, {the $Z_2$ symmetry gets spontaneously broken and the domain walls are formed with white noise fluctuation. }
 
\paragraph{$c_H\lesssim 0.1$}
This case is our primary focus. Note that we still assume that $m_{\rm Hubble}^2\gg m_0^2$, and thus, we have a symmetric phase of $\phi$ during inflation. However, 
\begin{equation}
m_{\rm Hubble}^2 \;{\lesssim}  \;H_{\rm inf}^2,
\end{equation}
implies that inflationary fluctuations of $\phi$ are not suppressed at superhorizon scales, and thus $\phi$ in each Hubble horizon diffuses like the random walk during inflation. 
The diffusion due to the random walk leads to the growth of the variance, $\phi^2\sim t H_{\rm inf}^3/(2\pi)^2$, which is balanced by the classical motion due to the mass term, $m_{\rm Hubble}^2 \phi^2/2$. If the inflation lasts long enough, we obtain the equilibrium distribution with the probability~\cite{Starobinsky:1986fx,Starobinsky:1994bd,Hardwick:2017fjo, Graham:2018jyp,Guth:2018hsa,Alonso-Alvarez:2019ixv, Nakagawa:2020eeg,Murai:2023xjn}\footnote{This stochastic distribution of scalar field was discussed in the context of the axion dark matter~\cite{Graham:2018jyp,Guth:2018hsa,Ho:2019ayl,Matsui:2020wfx, Nakagawa:2020eeg,Murai:2023xjn} (without the non-minimal coupling), relaxion~\cite{Chatrchyan:2022pcb}, and scalar/vector boson dark matter with the non-minimal coupling~\cite{Alonso-Alvarez:2019ixv}.}
\begin{equation}
\label{eq:dist}
    P[\phi]\propto \exp{-\frac{8\pi^2}{3H_{\rm inf}^4}{\left(V(\phi)+\frac{m_{\rm Hubble}^2 \phi^2}{2}\right)}}.
\end{equation}
This is a probability distribution $\phi$ in a single Hubble patch during inflation. 
By neglecting the contribution of $V$, this distribution is reached with a time scale
\begin{equation}
    N_{\rm eq}\sim \frac{H_{\rm inf}^2}{m^2_{\rm Hubble}}\sim \frac{1}{6c_H}.
\end{equation}
The scalar potential $V(\phi)$, especially the quartic coupling, could reduce the amount of e-folds required to reach the equilibrium.\footnote{As discussed in the main text, the initial fluctuations for domain walls transition from inflationary to thermal in this case.}
If $N_{\rm eq}$ is much smaller than the e-folds for the observable Universe, $N$,  
$\phi$ in the whole {observable} Universe follows this distribution, i.e.  the population bias is close to zero if $N\gg N_{\rm eq}$. This is the case with
\begin{equation}
    c_H={\cal O}(0.001-0.1).
\end{equation} 
The effect of the finite mass slightly deviates the power spectrum of the initial fluctuations from the scale-invariance, but our conclusions do not change.

The above discussion shows that unless $c_H(>0)$ is much smaller than unity (i.e. $c_H < 10^{-3}$), domain walls are naturally formed with either the initial condition of white noise or scale-invariant fluctuations depending on the size of $c_H$.

\subsection{Case of axionic domain walls}
Next, we consider the axionic domain wall formation. As a UV theory, we consider a model with an approximate global $\mathop{\rm U}(1)$ symmetry. 
The potential is given similarly as 
\begin{equation}
V= -m^2_0| \Phi|^2+\frac{\lambda }{4}|\Phi|^4
\end{equation}
where the potential minimum is
\begin{equation}
\left\langle{|\Phi|}\right\rangle=\sqrt{\frac{m_0^2}{2\lambda}}\equiv \frac{f_a}{\sqrt{2}}
\end{equation}
and a Nambu-Goldstone boson or axion, $a$, appears in  the phase,
\begin{equation}
\Phi = \left(\frac{f_a}{\sqrt{2}}+s\right)e^{i a/f_a }.
\end{equation}
Here $s$ is the radial mode.  The axion acquires a potential via the explicit breaking of the $\mathop{\rm U}(1)$ symmetry, such as $\delta V\propto \Re \Phi.$

Again let us consider a non-minimal coupling  
\begin{equation}
{\cal L}= -c_H |\Phi|^2R,
\end{equation}
and the resultant Hubble-induced mass 
\begin{equation}
m_{\rm Hubble}^2= 6c_H H_{\rm inf}^2,
\end{equation}
for the $\Phi$ field.
Let us consider again the effect of this term by neglecting the bare potential of $\Phi$ during the inflation. 
\paragraph{$c_H\gtrsim 0.1$}
We have the symmetry restoration during inflation. Since $\Phi$ is heavy, the inflationary fluctuation is suppressed. 
After the inflation, when the tachyonic mass $-m_0^2$ takes over the Hubble-induced mass, the U(1) symmetry gets spontaneously broken. We obtain the usual string-wall network. 
This system has an initial fluctuation approximated by the white noise for $\Phi$.

\paragraph{$0\lesssim c_H\lesssim 0.1$}
In this case, we have the  equilibrium distribution of the two fields,
\begin{equation}
\label{eq:dist2}
    P[\Re \Phi, \Im \Phi]\propto \exp{- \frac{8\pi^2}{3 H_{\rm inf}^4}\left(V+{\frac{m_{\rm Hubble}^2 ((\Re \Phi)^2+(\Im \Phi)^2)}{2}}\right)}.
\end{equation}
{We have a flat distribution for the axion $\propto \arg \Phi $ by neglecting the axion mass} (The inflationary dynamics rarely change the axion potential because of {the $\mathop{\rm U}(1)$ symmetry}).  
Again, we obtain the field configuration leading to the formation of strings if $N>N_{\rm eq}\sim 1/(6c_H).$
However, this follows the string-wall network from inflationary fluctuation, which is not our focus.\footnote{We can avoid the string formation if we introduce a sufficiently large explicit breaking term of $\mathop{\rm U}(1)$ in non-minimal coupling, $\Phi^2 R^2$. We also point out it is important to take account of the long-range correlation when we discuss the string-wall network from the inflationary fluctuation. Our lattice simulation shows a longer lifetime against the population bias than the white noise case, and larger string loops, which can give interesting gravitational wave spectra. }

We, therefore, need to consider
\begin{equation}
    N\lesssim N_{\rm eq} \sim 1/(6 c_H).
\end{equation}
In this case, we expand $\Phi=\bar\Phi+\delta \Phi$, where $\bar \Phi$ follows the equilibrium distribution. 
By neglecting the mass effect, the $\delta \Phi$ distribution from the random jump forms a disk with a radius 
\begin{equation}
    {\sqrt{N} \frac{H_{\rm inf}}{2\pi}}\sim \sqrt{6 c_H N} {|\bar\Phi|/\xi}.
\end{equation}
Here we have introduced $\xi$ to characterize the tuning from the typical value of $\bar\Phi$, the $1\sigma$ deviation of the probability distribution, $|\bar{\Phi}|= \xi  \sqrt{\frac{3H_{\rm inf}^4}{4\pi^2 m_{\rm Hubble}^2}}$.  
If $c_H\sim {\cal O}(0.001)$, the domain wall without string can be formed with ${\cal O}(10\%)$ tuning, because the symmetric point cannot be covered by the disk with $\xi={\cal O}(1)$, while the axion potential hilltop can be covered at a probability of ${2N_{\rm DW}\times \sqrt{N} \frac{H_{\rm inf}}{2\pi}}/(2\pi \bar{\Phi})$ with $N_{\rm DW}$ being the domain wall number.

\paragraph{$c_H\lesssim 0$}
A much simpler scenario is the case of the negative Hubble-induced mass\cite{Takahashi:2020tqv}.
If the $\Phi$ has a negative mass-squared, it is driven to the non-zero  expectation value,
\begin{equation}
    |\Phi| \sim \sqrt{2|m_{\rm Hubble}|^2/\lambda}
\end{equation}
induced by the Hubble-induced mass. Here, we include the quartic coupling but neglect the $m_0^2$ term. 
This is of ${\cal O}( H_{\rm inf}) $ if
\begin{equation}
    {\lambda\sim |c_H|}.
\end{equation} 
Then it is very natural to have the domain wall formation without  strings because 
the random jump of the axion $a\sim \sqrt{N} H_{\rm inf}/2\pi$ is around the effective decay constant $f_{\rm eff} \simeq |\Phi| \sim H_{\rm inf}.$ 
If $|c_H|$ is greater than or around unity, the radial direction becomes heavier than $H_{\rm inf}$, and the fluctuation in the radial direction can be neglected. Even if this is not the case, the string formation is disfavored by the field probability distribution. After inflation, we need to keep $\Phi$ in the broken phase. To this end, we can consider the interaction of negative thermal mass squared through the portal coupling between $\Phi$ and the Standard Model Higgs field.

\section{Finite box size effect on the area parameter}
\label{app:2}

We have investigated the differences in the evolution of the area parameter between the cases with white noise and scale-invariant initial fluctuations in Sec.~\ref{subsec:evolution_of_the_area_param}. We have found that not only do the two cases show different values of the area parameter, but they also exhibit different growth rates, with the scale-invariant case showing slightly more rapid increase with time. See Figs.~\ref{fig:wall_length_int}, \ref{fig:wall_area} and \ref{fig:wall_area_finelat}. Here we show that the apparent growth of the area parameter is largely due to the finite box size effect, focusing on the 2D case, as this effect was already shown in Figs.~\ref{fig:wall_area} and \ref{fig:wall_area_finelat} for the 3D case.

Table~\ref{tab:A2_growth_rate} shows the results of growth rates with various box sizes $L_{\rm box}$ and the grid number $N_{\rm grid}$. The growth rate is the slope obtained by the linear fit\footnote{We choose the linear growth over the conformal time for simplicity.} of the area parameter versus time in the range of $\tau = 6-32/m_0$. The growth of the area parameter is more suppressed with larger box sizes and higher resolution.\footnote{This fact is also obtained in Ref.~\cite{Gonzalez}. 
In Ref.~\cite{Gonzalez}, the author also showed that, to be more precise, the growth rate depends on the infrared cutoff of the initial fluctuations.
This was demonstrated by investigating cases with a fixed box size but varying the infrared cutoff of the fluctuations, which turned out to be consistent with the simulation results obtained by varying the box size without the infrared cutoff.
} These larger-scale simulations indicate that the broader momentum space is covered. The mild dependence of the growth rate on the box size in the scale-invariant cases is plausible because the smaller area parameter can be attributed to the fluctuations with superhorizon-scale correlation and the number of the Hubble horizons contained in the box decreases with time.
Thus we consider that the mild growth of the area parameter found in the simulation is largely due to the finite box size, and there is no significant difference in the growth rate between the two different initial fluctuations.

\begin{table}[!t]
    \centering
    \begin{tabular}{cc|>{\centering\arraybackslash}p{2.0cm}|>{\centering\arraybackslash}p{2.0cm}}
        \hline\hline
        \multirow{2}{*}{$m_0 L_{\rm box}$} & \multirow{2}{*}{${N_{\rm grid}}$} & \multicolumn{2}{c}{growth rate ($10^{-3}m_0$)}\\
        & & WN & SI \\
        \hline
        $64\pi$  & $8192$  & $6.9$ & $14.5$     \\
        $64\pi$  & $16384$ & $6.4$ & $10.9$     \\
        $128\pi$ & $16384$ & $6.8$ & $9.3$      \\
        $128\pi$ & $32768$ & $5.9$ & $8.2$      \\
        $256\pi$ & $32768$ & $6.6$ & $6.3$      \\
        \hline\hline
    \end{tabular}
    \caption{
    The increasing rate of the area parameter ${\cal A}_2$ which is evaluated by the linear fit of the data in the range of $\tau = 6 - 32 / m_0$.
    For each case, the data are obtained by averaging the results of 20 realizations.
    The growth is less pronounced with the larger box and higher resolution.
    }
    \label{tab:A2_growth_rate}
\end{table}

\bibliography{ref}
\end{document}